\documentclass[journal=apchd5,manuscript=article]{achemso}

\usepackage{chemformula} % Formula subscripts using \ch{}
\usepackage[T1]{fontenc} % Use modern font encodings

\author{Maximilian Rödel}
\affiliation{Lehrstuhl für Experimentelle Physik VI, Universität Würzburg, Am Hubland, 97074 Würzburg, Germany}
\email{maximilian.roedel@uni-wuerzburg.de}

\author{Luca Nils Philipp}
\affiliation{Institut für Physikalische und Theoretische Chemie, Universität Würzburg, Am Hubland, 97074 Würzburg, Germany}

\author{Jin Hong Kim}
\affiliation{Center for Nanosystems Chemistry (CNC), Universität Würzburg, Theodor-Boveri-Weg, 97074 Würzburg, Germany}

\author{Matthias Lehmann}
\affiliation{Center for Nanosystems Chemistry (CNC), Universität Würzburg, Theodor-Boveri-Weg, 97074 Würzburg, Germany}
\alsoaffiliation{Institut für Organische Chemie, Universität Würzburg, Am Hubland, 97074 Würzburg, Germany}

\author{Matthias Stolte}
\affiliation{Center for Nanosystems Chemistry (CNC), Universität Würzburg, Theodor-Boveri-Weg, 97074 Würzburg, Germany}
\alsoaffiliation{Institut für Organische Chemie, Universität Würzburg, Am Hubland, 97074 Würzburg, Germany}

\author{Roland Mitric}
\affiliation{Institut für Physikalische und Theoretische Chemie, Universität Würzburg, Am Hubland, 97074 Würzburg, Germany}
\email{roland.mitric@uni-wuerzburg.de}

\author{Frank Würthner}
\affiliation{Center for Nanosystems Chemistry (CNC), Universität Würzburg, Theodor-Boveri-Weg, 97074 Würzburg, Germany}
\alsoaffiliation{Institut für Organische Chemie, Universität Würzburg, Am Hubland, 97074 Würzburg, Germany}
\email{wuerthner@uni-wuerzburg.de}

\author{Jens Pflaum}
\affiliation{Lehrstuhl für Experimentelle Physik VI, Universität Würzburg, Am Hubland, 97074 Würzburg, Germany}
\alsoaffiliation{Center for Applied Energy Research e.V. (CAE Bayern), Magdalene-Schoch-Str. 3, 97074 Würzburg, Germany}
\email{jpflaum@physik.uni-wuerzburg.de}

\title[Article Topic]{Anisotropic Photo-Physical Properties of Plexcitons in Strongly Coupled Metal-Organic Thin Films}

\abbreviations{}

\begin{document}

%%%%%%%%%%%%%%%%%%%%%%%%%%%%%%%%%%%%%%%%%%%%%%%%%%%%%%%%%%%%%%%%%%%%%
%% The "tocentry" environment can be used to create an entry for the
%% graphical table of contents. It is given here as some journals
%% require that it is printed as part of the abstract page. It will
%% be automatically moved as appropriate.
%%%%%%%%%%%%%%%%%%%%%%%%%%%%%%%%%%%%%%%%%%%%%%%%%%%%%%%%%%%%%%%%%%%%%

%%%%%%%%%%%%%%%%%%%%%%%%%%%%%%%%%%%%%%%%%%%%%%%%%%%%%%%%%%%%%%%%%%%%%
%% The abstract environment will automatically gobble the contents
%% if an abstract is not used by the target journal.
%%%%%%%%%%%%%%%%%%%%%%%%%%%%%%%%%%%%%%%%%%%%%%%%%%%%%%%%%%%%%%%%%%%%%
\begin{abstract}
Exciton plasmon polaritons have gained increasing interests over recent years due to their versatile properties emerging by the underlying light-matter coupling and making them potential candidates for new photonic applications. We have advanced this concept by studying thin films of laterally aligned J-type aggregates of self-assembled tetra-bay phenoxy-dendronized perylene bisimide (PBI) molecules, arranged in a helical manner of three strains on a silver surface. As a result of the interaction between the uniformly aligned dipole moments and the surface plasmons of a thin silver layer underneath, the excitonic state at 1.94\,eV evolves into dispersions in absorption and emission, both characterized by a distinct anisotropy. The coupling constant defined by the scalar product of the transition dipole moment $\vec{\mu}$ and the surface plasmon wavevector $\vec{k}_x$ shows a pronounced two-fold rotational symmetry with values between almost 0 to 28\,meV. Complementary TD-DFT calculations of the angular dependent absorption and photoluminescence provide insights in the coherent energy exchange between the excitonic and plasmonic sub-systems. Additionally, power dependent PL studies yield first evidence that the diffusion length of the coupled exciton-plasmon polaritons exceeds that of the mere Frenkel state in neat PBI by at least one order of magnitude. Our results not only demonstrate the possibility to control the photo-physical properties of strongly coupled states by their spatially anisotropic light-matter interaction but also reveal innovative strategies to influence opto-electronic device operation by the directional transport of hybrid state energy.
\end{abstract}

%%%%%%%%%%%%%%%%%%%%%%%%%%%%%%%%%%%%%%%%%%%%%%%%%%%%%%%%%%%%%%%%%%%%%
%% Start the main part of the manuscript here.
%%%%%%%%%%%%%%%%%%%%%%%%%%%%%%%%%%%%%%%%%%%%%%%%%%%%%%%%%%%%%%%%%%%%%
\section{Introduction}
Strong light-matter coupling constitutes a well-established phenomenon in today’s quantum-electrodynamics and has been applied to extend the photo-physical properties of functional materials\cite{nizar2021emergent} or to steer photochemical reactions\cite{garcia2021manipulating}. Besides light-matter coupling in specially designed cavities \cite{kolb2017hybrid, tropf2017influence, betzold2018tunable, dusel2017three}, an alternative approach relies on the strong interaction between the collective excitation of a quasi-free electron gas (plasmon) in a metal and the two-particle electron-hole excitation (exciton) in a semiconductor. These mixed states, also referred to as plexcitons, exhibit features of both, the plasmonic as well as the particle-like excitonic system and, as a result, offer a variety of new phenomena such as quasi-Bose-Einstein condensation or polariton lasing \cite{kasprzak2006bose, schneider2013electrically}. The anti-crossing (AC) emerging by the interaction of the related dispersions and lifting the degeneracy between the upper and lower plexciton branch marks a striking feature of such coupled systems. Despite of localized surface plasmons polaritons (LSPP) are mostly used to generate light-matter coupling \cite{ajaykumar2023single}\,, we here choose a different approach by coupling the excited states in molecular thin films to the surface plasmon polaritons (SPP) of an adjacent metallic layer. Such hybrid stacks constitute a versatile platform to study angular dependent static photo-physical properties such as the absorption and emission characteristics, but, in addition, allow also for analysis of the transport behavior of photo-excited hybrid states and its correlation to the anisotropic in-plane coupling between localized (Frenkel-type excitons) and delocalized (plasmons) excitations in strongly coupled metal-organic structures.\\
For this purpose, we chose self-assembled J-type aggregates of a tetra-bay phenoxy-dendronized perylene bisimide (PBI) \textbf{1}\cite{kaiser2009fluorescent, merdasa2014single, lin2010collective, tian2012reorganization} in Figure \ref{Fig.: Experimental and Methods}a), a highly stable organic semiconductor with considerable oscillator strength and, thus, high photon-yield\cite{muller2022photon}, which by its liquid–crystalline (LC) phase as a triple-stranded helical arrangement (see SI’s) offers an oriented, well-ordered template. For the metal thin film providing the SPP our choice of material was silver as it matches the spectral characteristics of the molecular layer close-by and shows low damping of collective electronic excitations by the small imaginary part of its dielectric constant. Coupling the photo-excited states in the PBI \textbf{1} layer to the SPP in the silver layer underneath thus generates a lateral azimuthal angle ($\varphi$)-dependent anisotropy in the emerging plexciton dispersions. Besides evidencing and analyzing these anisotropy effects by their signature in the measured plexciton dispersions, complementary theoretical calculations provide crucial insights into the photo-physical nature of these light-matter excitations and disclose strategies towards manipulating and adjusting their properties.\\
The Hamiltonian for strongly coupled exciton plasmon polaritons $ H_{\vec{k}} ^N = H_{SPP} + H_{Exc} + H_{C}$ can be deduced by the adapted Jaynes Cumming model\cite{jaynes1963comparison} as described in our previous study\cite{rodel2022role} with $H_{SPP}$  being the operator describing the plasmonic excitations, $H_{Exc}$  the excitonic ones and $H_C$  the excitations resulting by the coupling between both\cite{jaynes1963comparison}. After deriving the Hamilton operator in matrix representation, the energy eigenvalues of the new emerging upper and lower plexciton branches can be calculated by the characteristic determinant according to 
\begin{equation}
E_\pm= \frac{1}{2} (E_{SPP}+E_{Exc}) \pm \frac{1}{2} \sqrt{(E_{SPP}-E_{Exc})^2+4V^2}.
\label{eq.: Plexciton Branches}
\end{equation}
The coupling constant $V=g_{\vec{\mu}} ^N (\vec{k})= \sqrt{n \int_{z_0}^{z_0+d} |{g_{\vec{\mu}}(\vec{k};z)}|^2 dz}$ contains the contributions of all $N$ emitters located at the distance $z_0$ within a given layer of thickness $d = (z_0 + d) - z_0 $ multiplied by the volume density $n=\frac{N_s N_L}{A d}$. Here $N_S$ denotes the number of quantum emitters per layer, $N_L$ the number of layers and $A$ the projected area of the emissive layer on the metal-dielectric interface. More precisely,
\begin{equation}
g_{\vec{\mu}}(\vec{k};z) = \sqrt{\frac{\omega(\vec{k})}{2 \epsilon_0 L(\vec{k})}} e^{-k_z z} \vec{\mu} \cdot \vec{E}(\vec{k})
\label{eq.: Plexciton Coupling Strength}
\end{equation}
describes the coupling of the transition dipole moment $\vec{\mu}$ of a distinct molecule or aggregate at normal position $z$ to the electric field component $\vec{E}(\vec{k})=\left( \hat{e}_{\vec{k}} + i \frac{|{\vec{k}|}}{k_z} \hat{e}_z \right)$ of the SPP. By the evanescent nature of the out-of-plane component of the surface plasmon polariton field, the coupling strength decays exponentially and, therefore, leads to a strong dependence along $z$-direction, while $L(\vec{k})$ is the effective in-plane length of the plasmon mode. 
\section{Materials and Methods}
\begin{figure}
\includegraphics[]{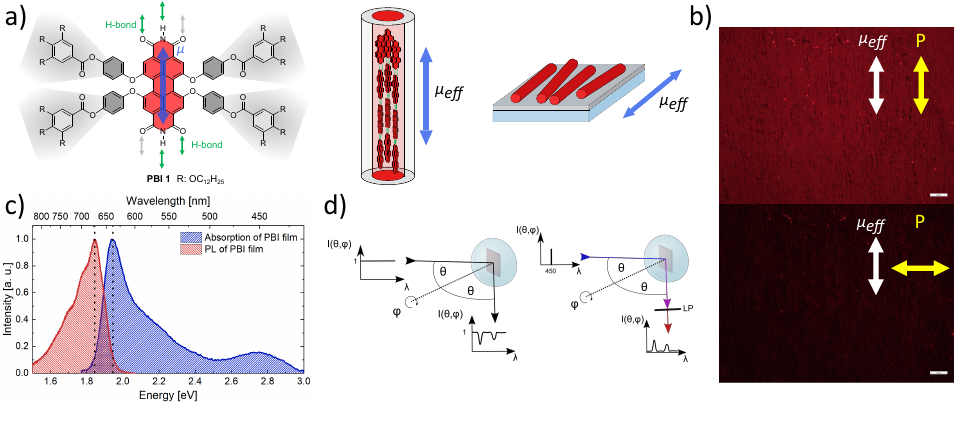}
\centering
\caption{a) Chemical and supramolecular structure of the tetra-bay phenoxy-dendronized PBI \textbf{1} and the resultant self-assembled columnar liquid-crystalline thin film used as excitonic material in this study. While off-center spin-coating (OCSC) the thin active layer on top of the substrate, aligned J-aggregates are forming via complementary hydrogen bonding as well as $\pi-\pi$-stacking in a triple-stranded helical columnar structure. Doing so, the individual electrical transition dipole moments $\vec{\mu}$ (blue) form to an overall effective moment $\vec{\mu}_{eff}$ oriented along the helical axis. Utilizing this effect leads to laterally aligned cylinders with pronounced in-plane anisotropic structural as well as optical properties. b) Polarization dependent fluorescence microscope images of an PBI J-aggregate film deposited on silver (excitation: 530-580\,nm; detection: >600\,nm). The arrows indicate the orientation of the effective transition dipole moment within the film plane and thus, the polarization direction of excitation as well as the orientation of the analyser for detection. c) Absorption and emission of an OCSC deposited PBI thin film under normal incidence conditions. d) Scheme of the Kretschmann geometries for measuring the spectral resolved plexciton absorption (left) and photoluminescence (right). In case of the latter, the white light source is replaced by a blue laser of 450\,nm wavelength together with a long pass (LP) filter in the detection path.}
\label{Fig.: Experimental and Methods}
\end{figure}
For our studies we used the tetra-bay phenoxy-dendronized PBI \textbf{1} with free imide protons for complementary hydrogen bonding displayed in Figure \ref{Fig.: Experimental and Methods}a). The supramolecular building block has four bay substituents with each three solubilizing alkyl chains (grey)  which are connected to the main PBI chromophore $\pi$-core (red)\cite{herbst2018self}. The volumetric ratio between $\pi$-core and peripheral alkyl substituents determines the strand number of the columnar liquid-crystalline phase that is formed upon hydrogen bond-directed self-assembly of the PBI units\cite{hecht2020supramolecularly}. The corresponding transition dipole moment $\mu$ of the $S_0-S_1$ monomer is indicated with a blue arrow and aligned along the long molecular N,N’ axis of the PBI building block as well as the macroscopic axis of the columnar mesophase. Next to the molecular structure this spatial arrangement of the single molecules within the LC phase is depicted. Within the mesophase, by arranging in a slip-stacked fashion via strong directional hydrogen bonds as well as $\pi-\pi$-stacking, the PBIs self-assemble into a triple-stranded helical column displaying pronounced J-type exciton coupling with absorption maximum at $\lambda_{max} \approx 639\,nm$ (Figure \ref{Fig.: Experimental and Methods}c), blue)\cite{herbst2018self}. By this directional alignment via supramolecular interactions between the PBI chromophores within the helical strands, the resultant transition dipole moment of the J-aggregate within one LC column is also effectively aligned and can be seen as one effective transition dipole moment $\mu_{eff}$ orientated along the column axis (Figure \ref{Fig.: Experimental and Methods}a)). By utilizing off-centered spin coating (OCSC) (see SI’s; Figure S1)\cite{kim2022wavelength, yuan2014ultra} and the fact that the strands are collectively aligned by the centrifugal force, we systematically achieved an overall effective orientation of the whole LC layer. To prepare the metallic layer that provides the propagating surface plasmons, 2\,nm of chromium and 45\,nm of silver were vacuum deposited (base pressure of $10^{-7}$\,mbar) on pre-cleaned glass substrates. Afterwards the PBI layer was processed on top via OCSC. We fabricated, by default, two kinds of samples: one reference sample on a bare glass substrate with aligned LC columns for reference/characterisation measurements and another sample as glass/Cr/Ag/PBI \textbf{1} J-aggregate for the plexciton measurements as described before.\\
Figure \ref{Fig.: Experimental and Methods}b) displays a polarization-dependent fluorescence microscope image of the fabricated PBI \textbf{1} thin film on silver upon irradiation by green light (excitation: 530-580\,nm; detection: >600\,nm). While the white arrows indicate the effective transition dipole moment $\mu_{eff}$ of the organic layer and, thus, the polarization of the emitted light, the yellow arrow (P) designates the orientation of the analyser in the detection pathway. As can be seen by the intensity contrast, there exists a pronounced anisotropy in polarization caused by the preferential orientation of the transition dipole moments within the sample with a dichroic ratio ($D_\lambda$) of about $3.37 \pm 0.57$ at 639\,nm. Additionally, we optically characterized in Figure \ref{Fig.: Experimental and Methods}c) a reference sample on glass without an intermediate silver layer by photoluminescence (PL; red) and UV-Vis spectroscopy (absorption; blue). The vertical lines indicate the main $S_1-S_0$ transition at 1.850\,eV for PL and at 1.949\,eV ($S_0-S_1$) for absorption, respectively. We utilized the high energy absorption band of the $S_0-S_2$ transition of the PBI \textbf{1} LC thin film at approximately 2.75\,eV ($\approx450$\,nm) for photo-excitation in our PL and plexciton emission experiments (see SI's; Figure S3). In Figure \ref{Fig.: Experimental and Methods}d) the two measurement configurations in Kretschmann geometry are illustrated which we applied in our plexciton dispersion studies. On the left side, utilizing a white light source, the scheme to measure the angular dependent absorption is shown. Replacing this source by a monochromatic blue laser of 450\,nm wavelength and an additional long pass (LP) filter to suppress the reflected laser line in the detection path, the plexciton photo-luminescence was examined as illustrated on the right in Figure \ref{Fig.: Experimental and Methods}d).\\
\section{Plexciton Absorption Dispersions}
\begin{figure}
\includegraphics[]{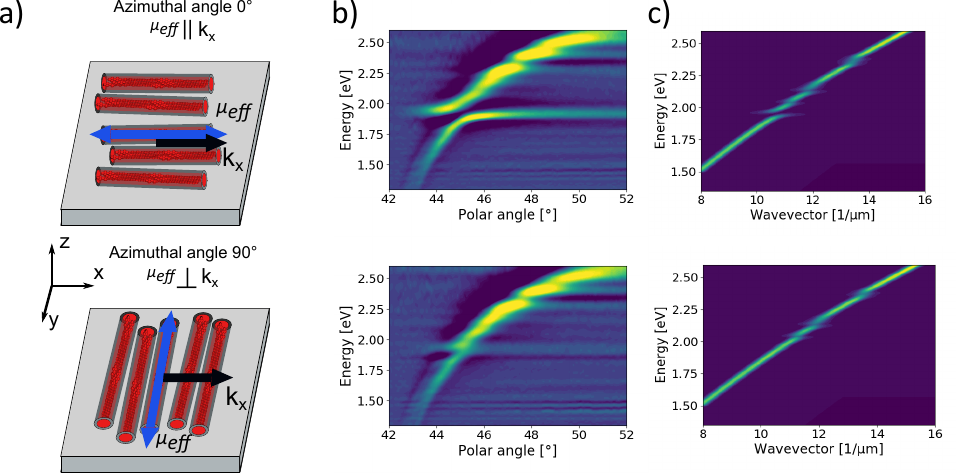}
\centering
\caption{a) Relative sample orientation for the anisotropic plexciton dispersions measured for two limiting configurations. The perpendicular alignment between the PBI aggregate transition dipole moment $\vec{\mu}_{eff}$ and the wavevector $\vec{k}_x$ of the incident light, i.e. of the plasmonic wave, refers to the situation of no coupling (lower column), whereas the parallel orientation of $\vec{\mu}_{eff}$ and $\vec{k}_x$ corresponds to the case of strong light-matter coupling (upper column). b) Measured and c) calculated plexciton dispersion for the two limiting configurations.}
\label{Fig.: Absorption Plexciton Dispersion}
\end{figure}
\begin{figure}
\includegraphics[]{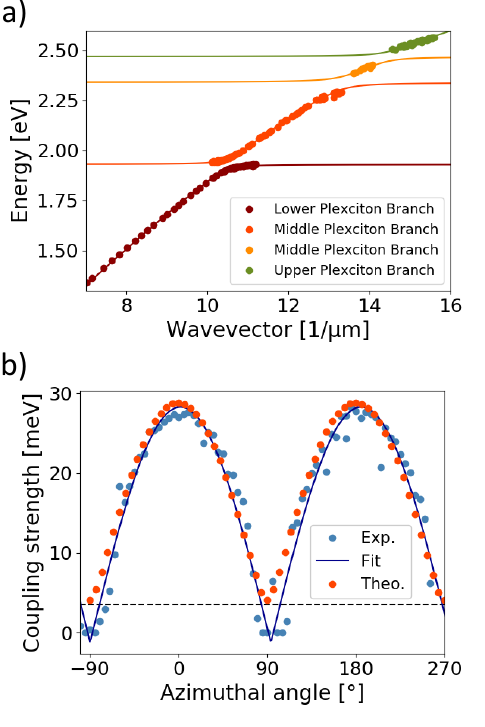}
\centering
\caption{a) Plexciton dispersion in the strong coupling regime for $\varphi=0^\circ$. Three distinct anti-crossings can be identified. b) Coupling strength as function of azimuthal angle, determined for the anti-crossing at 1.93\,eV. The blue line is a fit based on the absolute cosine of the angle between the electric field of the surface plasmon, thus the direction of the wavevector $\vec{k}_x$ and $\vec{\mu}_{eff}$ while the orange dots display the calculated values of the coupling strength. The dashed horizontal line at 3.5\,meV indicates the estimated lower resolution limit of the coupling strength in the measurement.}
\label{Fig.: Result Absorption Plexciton Dispersion}
\end{figure}

The plexciton measurements were carried out with an optical goniometer setup working in modified Kretschmann geometry, presented in detail in a previous study.\cite{rodel2022role} Utilizing a half sphere lens we are able to access the azimuthal angle $\varphi$ as additional degree of freedom, sketched on the left in Figure \ref{Fig.: Experimental and Methods}d). This setup allows to rotate the whole sample in-plane while measuring on the same spot, thus enabling to record the 2D plexciton dispersion of our Ag/PBI \textbf{1} hybrid stack at each polar angle $\theta$.\\
In Figure \ref{Fig.: Absorption Plexciton Dispersion} representative experimental and theoretical dispersion curves for the two limiting cases of weakest and strongest coupling are shown, respectively. Here, we displayed the second derivative of the normalized data to increase its visibility. Accordingly, the yellow areas refer to minima in the reflection spectra and, thus, correspond to the excitation of a plexciton by absorbing a photon with matching energy and momentum. Fitting these minima by a Lorentzian oscillator model for each branch one can calculate the plexciton energy for a given wavevector as depicted by the dots in Figure \ref{Fig.: Result Absorption Plexciton Dispersion}a) for the case of strongest coupling ($\varphi = 0^\circ$). Modelling, in a next step, the plexciton branches according to Equation \ref{eq.: Plexciton Branches} (lines in Figure \ref{Fig.: Result Absorption Plexciton Dispersion}a)) we can deduce the energetic positions and splittings, i.e. coupling strengths, of the detected anti-crossings, the values of which are listed in Table \ref{tbl: Absorption Plexciton Fit Values}.\\
\begin{table}
  \caption{Energetic position $E_{Exc}$ and related coupling strength $V$ of the three anti-crossings (AC) deduced from the plexciton dispersion curves in Figure \ref{Fig.: Result Absorption Plexciton Dispersion}a).}
  \label{tbl: Absorption Plexciton Fit Values}
  \begin{tabular}{|c|c|c|}
    \hline
    AC-Nr.  & $E_{Exc}$ [eV] & $V$ [meV] \\
    \hline
    \hline
    1 & 2.47 & 27 \\
    \hline
    2 & 2.34 & 33 \\
    \hline
    3 & 1.93 & 0-28 \\
    \hline
  \end{tabular}
\end{table}For the case of strongest coupling, i.e. at $\varphi = 0^\circ$, we can identify three distinct anti-crossings in Figure \ref{Fig.: Result Absorption Plexciton Dispersion}a). The two high-energy anti-crossings at 2.47\,eV and 2.34\,eV do not show any significant dependence on the azimuthal angle, whereas the anti-crossing at 1.93\,eV, related to the J-aggregates’ absorption maximum in Figure \ref{Fig.: Experimental and Methods}c), almost vanishes when reaching $\varphi = 90^\circ$.\\
To further investigate the plexcitonic dispersions and the anisotropy of the energetically lowest anti-crossing, we applied our model for the description of strong coupling between excitons and SPP modes to the system at hand (for details of the calculation see SI's). This allowed us to calculate the mean plexcitonic dispersions for the two representative plexciton dispersions, which are displayed in Figure \ref{Fig.: Absorption Plexciton Dispersion}c). In accordance with the experimental dispersions, the simulated ones show two large anti-crossings at 1.93\,eV and between 2.3\,eV and 2.4\,eV. Remarkably, they also exhibit some very faint signatures of anti-crossings between 2.0\,eV and 2.1\,eV, which is in agreement with the experimental data for the weak and strong coupling case. However, the energetically highest anti-crossing, which is located at an energy of 2.47\,eV in the experimental dispersions, is absent in our simulations.\\
We attribute this anti-crossing to uncoupled PBI monomers within the organic layer. As we have already observed in previous studies on phthalocyanine monomers at gold interfaces [5], even a small fraction of monomers can have a significant impact on the plexcitonic signal, since the strength of the signal is strongly dependent on the distance to the silver surface. These monomers originate by the fact that under real conditions not all molecules will be incorporated during PBI J-aggregate formation, especially those located in close proximity to the substrate interface and interacting strongly with the underlying silver layer. Hence, we expect these uncoupled monomers to lead to a detectable plexciton signal despite their overall small fraction.\\
In accordance with the experimental dispersions, our calculations predict an anti-crossing between 2.3\,eV and 2.4\,eV. However, the energetic position of this anti-crossing can vary if the structure is rotated around the helical axis, because there are two different contributing excited states (for details see SI's Figure S8). The respective excitation energies of these states are 2.359\,eV and 2.386\,eV. Since there is no preferred alignment direction for the helix with respect to the plane perpendicular to the helical axis, we expect that the anti-crossing appears at the mean value of both excitation energies. Furthermore, the calculations confirm that the size of the splitting of this anti-crossing does not strongly depend on the azimuthal angle, i.e., the inner product $\vec{\mu}\cdot \vec{E}$ remains almost constant if one of the vectors is rotated with respect to the axis perpendicular to the silver surface. This can be rationalized by considering that for both states the coupling constant with the SPP is dominated by the contribution of the transition dipole moment, which is perpendicular to the silver surface and thus, does not change under rotations around the same axis. It should be mentioned here, however, that in the experiments the splitting at 2.34\,eV can show slight modulations (see reference measurement in Figure S2) due to a systematic uncertainty. Nevertheless, by our comprehensive data sets we are convinced that those features only weakly shift the anti-crossing position but not the overall coupling strength.\\
In contrast to the former anti-crossing, the plexcitonic dispersions confirm that the splitting of the energetically lowest anti-crossing at 1.93eV depends for the theory and experiment strongly on the orientation between the laterally aligned, long-ranged ordered PBI strands\cite{kaiser2009fluorescent, tian2012reorganization} and the electromagnetic field of the SPP. In this case, the calculations show that the component of the transition dipole moment in the direction of the helical strand determines the strength of the coupling. Consequently, this leads to an azimuthal angle dependent increase in the splitting of the anti-crossing due to the inner product $\vec{\mu} \cdot \vec{E} = |\vec{\mu}| |\vec{E}| cos(\varphi)$ if the SPP propagates in the direction of the transition dipole moments within the helical strand ($\varphi =0^\circ$), compared to the SPP propagating in the direction perpendicular to it ($\varphi =90^\circ$), as can be seen in Figure \ref{Fig.: Absorption Plexciton Dispersion}.\\
The experimentally determined coupling strengths are plotted together with theoretically calculated ones as function of the azimuthal angle in Figure \ref{Fig.: Result Absorption Plexciton Dispersion}b) for the complete data set (in addition, the complete sets of absorption dispersion data for each azimuthal angle visualized in the supplementary GIF). According to Equation \ref{eq.: Plexciton Coupling Strength}, the data was fitted by an absolute cosine indicated by the dark blue line in the graph. A maximum coupling strength of $(29.5\pm0.9)$\,meV at $\varphi = 0^\circ$ with a $180^\circ$ modulation in accordance with the in-plane alignment of the PBI helix strands could be deduced. We would like to point out that fitting the plexciton dispersions for very small coupling constants turns out to be challenging, due to the spectral overlap of the related absorption minima, i.e. their splitting, in the respective dispersion curves. We therefore rationalized a lower limit for the coupling strength of 3.5\,meV indicated by the dashed line in Figure \ref{Fig.: Result Absorption Plexciton Dispersion}b). The theoretical values shown as orange dots in Figure \ref{Fig.: Result Absorption Plexciton Dispersion}b) agree within the error with the experimental data from our optical measurements and therefore support the analysis carried-out.\\
\section{Plexciton Photoluminescence Dispersions}
\begin{figure}
\includegraphics[]{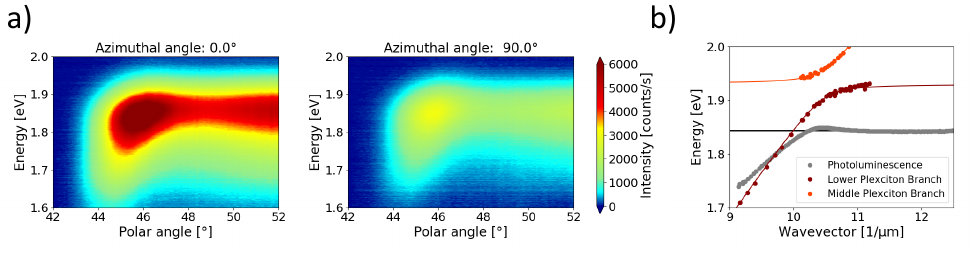}
\centering
\caption{a) Measurements of the plexciton emission at an azimuthal angle of $0^\circ$ (strongest coupling) and $90^\circ$ (weakest coupling). As for the absorption (see Figure \ref{Fig.: Absorption Plexciton Dispersion}), the photoluminescence intensity strongly depends on the respective coupling strength in the Ag/PBI hybrid films. The intensity scale bar refers to both Figures. b) Comparative presentation of the plexciton absorption (lower and middle plexciton branch in red and orange) and the photoluminescence (in grey) in the strong coupling regime as a function of the wavevector. The observed intersection between PL dispersion and absorption of the lower plexciton branch indicates a change in the dielectric landscape close to the metal interface upon changing the plasmonic part of the plexcitonic state.}
\label{Fig.: Emission Plexciton Dispersion}
\end{figure}

\begin{figure}
\includegraphics[]{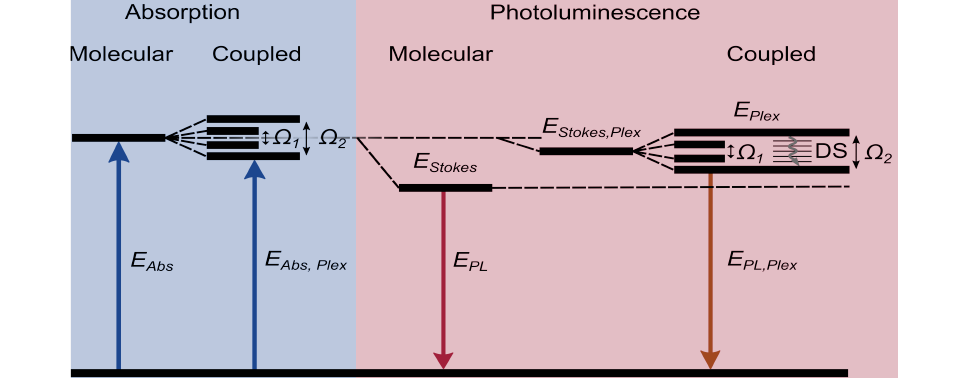}
\centering
\caption{Energy model for the absorption (blue shaded area) and photoluminescence paths (red shaded area). Upon coupling the molecular aggregate to the surface plasmon polariton the energetic levels split into two new branches. Since the plexciton possesses characteristics of both states, the Stokes shift is altered for the coupled system. Taking the additionally energetic shift of the energetic splitting into account, results in an overall energetic PL-shift of the coupled plexciton state. Higher energetic states relax via intermediated dark states (DS) into the lowest radiative state.}
\label{Fig.: Energetic Level Scheme}
\end{figure}

\begin{figure}
\includegraphics[]{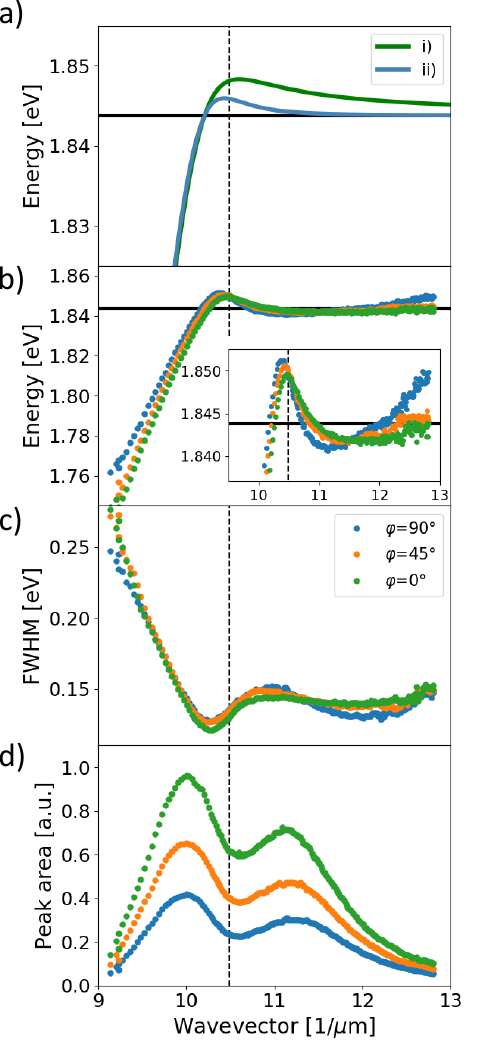}
\centering
\caption{a) Theoretical calculated PL-plexciton dispersion for the case i) and ii) described in the main text, corresponding to the strong and weak coupling, respectively. b) Measured PL-plexciton dispersion, calculated from the fitted peak maxima c) FWHM of the fitted peaks and d) Peak area/intensity of the fitted peaks for three different azimuthal angles $\varphi=0^\circ$ $(\vec{k}\parallel\vec{\mu})$, $45^\circ$, $90^\circ$ $(\vec{k}\perp\vec{\mu})$, in green, orange, and blue.}
\label{Fig.: Result Emission Plexciton Dispersion}
\end{figure}

While the anisotropic behavior of light-matter coupling can be considered remarkable itself and, to our knowledge, has not yet been reported for supra-molecular excitonic materials, the ability to specifically tune the coupling strength by the choice of azimuthal angle offers great promises in future opto-electronic implementation. As a stringent extension of our studies, we therefore have investigated the photoluminescence characteristics of these metal-organic hybrid states. For this purpose, the white light source for the absorption measurements was exchanged by a p-polarized cw-laser operating at a wavelength of 450\,nm and an excitation power of 1.65\,mW at the sample position (illustrated in Figure \ref{Fig.: Experimental and Methods}d), right). According to the spot diameter of $\leq 1$\,mm this results in a power density of $0.21 \frac{\text{W}}{\text{cm}^2}$.\\
In Figure \ref{Fig.: Emission Plexciton Dispersion}a) two representative plexciton emission spectra are plotted for the case of strongest and weakest coupling at $\varphi=0^\circ$ and $\varphi=90^\circ$, respectively (the complete data set is contained in the SI and GIF) with the intensity scale bar in counts/s referring to both Figures. Owing to the similar PL intensity behaviour in terms of energy and angular dependence, we attribute the observed emission exclusively to the lower plexciton branch of the anti-crossing at 1.93\,eV. Similar to Kasha’s rule plexcitonic states at higher energies are non-radiatively relaxing towards the lowest plexciton state via intermediated states. The upper plexciton branch of the lowest plexcitonic state decays into dark states (DS) and are able to radiatively relax from there into the ground state, as seen Figure \ref{Fig.: Energetic Level Scheme}. We have to emphasize that these states are only DS with respect to the plexcitonic state and, under certain circumstances, can still emit light. To independently test our hypothesis of exited/emitting plexcitons, we measured the angular dependent PL emission of the sample under s-polarized photo-excitation (see supplement and GIF) and, concordantly, the emission intensity almost vanishes and shows a significantly smaller PL-intensity compared to p-polarized excitation. We assume the mixed plasmon-exciton composition of the plexcitonic state to prevent a complete PL extinction in case of s-polarized excitation, together with a minor contribution by the non-perfect p- to s-conversion of our achromatic $\lambda$/2-plate.\\
Concordantly to the evaluation of the absorption above, we fitted the PL data with a Lorentzian function for each branch. Since we only detect a single, presumably the lowest emission branch for all azimuthal angles only one Lorentzian is needed for modelling (see Figure S6). The PL plexciton dispersion for $\varphi=0^\circ$ is plotted together with the corresponding absorption dispersion at $\varphi=0^\circ$ in Figure \ref{Fig.: Emission Plexciton Dispersion}b). As a striking feature, not to be expected at first glance, we observe an intersection of the lower absorption and emission branch at a defined wavevector of $k = 9.75 \dfrac{1}{\mu \text{m}}$. This remarkable feature could be explained by significant changes in the dielectric environment nearby the PBI J-aggregate during the excitation and emission process.\\
In literature, the plasmon dispersion in a three-layer model (metal substrate, dielectric spacer, organic film) is dependent on the dielectric constants $\epsilon_d$  and $\epsilon_{org}$ of the spacer and the organics, respectively\cite{yuen2016plexciton, yuen2018molecular}. This model can be adapted to our PL-plexciton sample geometry. However here, the third layer is originating not by a dielectric spacer, since such a layer is absent in our case, but corresponds to a fraction of photo-excited molecules within the organic layer. By generating a certain density of plexcitons with a defined plasmonic and excitonic part, depending on the angular detection and excitation, one can effectively alter the dielectric environment. Enhancing the plasmonic character, accompanied with an increase in delocalization of the plexcitonic state, thus, results in a change in polarizability and goes along with an effective change of the dielectric environment. This leads to a lifting of the PL plexciton branch with regard to its plasmonic character and thereby, to the crossing of both, the photoluminescence and the absorption dispersion.\\
A zoom-in of the Lorentzian-fitted PL dispersion is displayed in Figure \ref{Fig.: Result Emission Plexciton Dispersion}b) for three different azimuthal angles related to weakest ($\varphi=90^\circ$, blue dots), strongest ($\varphi=0^\circ$, green dots) as well as intermediate coupling strength at $\varphi=45^\circ$ (orange dots). Besides the qualitative similarity with the absorption dispersion, the PL emission maximum being prominent between $k = 10.25 - 10.50\dfrac{1}{\mu \text{m}}$ shows a distinct variation in energy and azimuthal angle. As such, with stronger coupling the peak maximum is shifted to lower energies by approximately 1-2\,meV while simultaneously, the correlated azimuthal angle, i.e. the wave vector, is increased. The associated decrease of the Stokes shift, i.e. the spacing between absorption and emission maxima, with increasing coupling strength suggests the following heuristic model depicted in Figure \ref{Fig.: Energetic Level Scheme}. In case of excitation by a photon of energy $E_{Abs}$, a molecule usually is transferred from its $S_0$ singlet ground state to an energetically higher lying vibrational state of its, e.g. $S_1$ manifold. From there, by ultra fast vibrational relaxation, the excited state relaxes to the zero vibrational state of the $S_1$ manifold where it decays under emission of a photon (PL) with distinct energy $E_{PL}$ back to its $S_0$ ground state for the next excitation cycle. The loss in energy due to the vibrational relaxation or dissipation can be detected by a red-shift in the emission compared to the absorption and is quantified by the Stokes shift, $E_{Stokes}$. In contrast to an isolated molecule within a matrix or molecules within an aggregate, the photo-dynamics in our strongly coupled exciton-plasmon polariton system is affected by additional states and transitions (s. scheme in Figure \ref{Fig.: Energetic Level Scheme}). In this case, the plasmonic part of the excitation supports a spatially more extended delocalization of the plexciton wavefunction (Wannier-type) compared to that of the localized, mere molecular electron-hole pair (Frenkel- or Charge-Transfer-type)\cite{quenzel2022plasmon}. Accordingly, this will alter the binding energy and, thus, reduces the Stokes shift $E_{Stokes,Plex}$ upon forming an plexcitonic state as shown in Figure \ref{Fig.: Energetic Level Scheme}. This leads, at first, to an overall blue-shift of the plexciton emission energy $E_{PL,Plex}$. However, the defined energetic splitting $\Omega$ of the newly arising plexciton state has to be considered. Since the emission into the ground state $S_0$ will proceed from the lowest energetic state this will counteract the initial blue-shift of $E_{Stokes,Plex}$ and will lead to the first contributions which alters the energetic position of the PL-plexciton emission dispersion.\\
This consideration needs be carried-out consistently for each possible $k$-value and will lead to the detected plexciton PL dispersion in Figure \ref{Fig.: Result Emission Plexciton Dispersion}b), depending on the plasmonic and excitonic character. At approximately $k = 10.49 \dfrac{1}{\mu \text{m}}$ (vertical dashed line) the energy of the PL emission shows a local maximum, which for all cases of coupling is higher than that of the neat J-aggregated PBI emission at 1.849\,eV measured by standard PL (horizontal bold line). We would like to emphasize, that this $k$-value also marks the position where the plasmonic and the excitonic character of the plexciton amounts to equal portions (50:50 mixture) as can be derived from the Hopfield coefficients displayed in the supplementary Figure S4.\\
For a more quantitative analysis of the photoluminescence energy, an additional theoretical model for the strong coupling case is proposed, which is based on the Hamiltonian for the strongly coupled exciton plasmon polariton and accounts for possible relaxation processes after the excitation into the lowest plexcitonic state (for a detailed description see SI's). The plexciton state which consists of a mixed excitonic and plasmonic part is composed of the two diabatic states $|0,e\rangle$ (a molecular state is excited, but no photon in the SPP mode) and $|1,g\rangle$ (no molecular state is excited, but a photon in the SPP mode) and thus has the two eigenstates $|+\rangle$ (upper plexciton branch) and $|-\rangle$ (lower plexciton branch). In general the plexcitonic state can now decay via the plasmonic part $|1,g\rangle$ or the excitonic part $|0,e\rangle$ into the molecular ground state $S_0$. Thus, the maximum of the photoluminescence energy at a given wavevector is a weighted average of the energies$ E_{SPP}(k)$ and $E_{Exc}$. The weights correspond to the probability, with which a plexcitonic state is excited and furthermore, relaxes into the respective state, i.e., purely excitonic or SPP state. While the lower plexcitonic state can now decay via the plasmonic part $|1,g\rangle$ or the excitonic part $|0,e\rangle$ into the molecular ground state $S_0$, for the upper branch it is assumed that it exclusively relaxes via the intermediated DS towards, but not completely into, the lower plexciton branch. These states form upon coupling of multiple helical strands and, as mentioned before, are only dark states with respect to the plexcitonic state. Therefore, these states can still radiatively relax into the ground state via a molecular excited state. Here, a distinction must be made as to whether the plexcitonic mixed state of the upper branch decays exclusively via its excitonic part (i) or whether both parts contribute equally (ii) to the decay into the dark states and thus, into the excited molecular state, resulting in a relaxation probability given by the Hopfield coefficients $|\langle -|0,e\rangle |^2 $ and $|\langle -|1,g\rangle |^2 $. We hence justify this assumption by the significantly higher density of states which favours the relaxation out of these states rather than out of the lower plexcitonic branch. For the calculation, we applied a linear fit to the experimental photoluminescence dispersion in the pure plasmonic region, i.e. where the deviation from the neat molecular excitation is high. The other parameters were chosen according to the experimental values.\\
The key difference in both cases is shown in Figure \ref{Fig.: Result Emission Plexciton Dispersion}a), where a clear wavevector shift of the PL-dispersion maxima can be seen. Despite its energetic offset, which marks the second controbution to the overall energetic Stokes-shift of the plexcitonic system, the curve for case ii) is dropping more sharply to the energetic ground state of the excited molecule (dashed horizontal line) than for case i), which is also seen in our experimental results. Even though the exact ratio of the two cases is not known in our studies, the following statement can be made: While the coupling is strong ($\Omega_2$) the main decay channel is weighted towards case i), while for weak coupling ($\Omega_1$) is weighted towards case ii) resulting in a shift of the energetic maxima towards lower $k$-values. This, however, cannot be the full explanation why, contradictory to the calculation, the energetic position of the PL-plexciton dispersion is shifting to higher values for the weak coupling case. For that, one has to take the combined energetic shift of the plexciton splitting and plexcitonic Stokes-shift as an additional factor into account, as discussed before and depicted in Figure \ref{Fig.: Energetic Level Scheme}. This not only compensates for the energetic blue shift caused by transitioning between both cases i) and ii), but, moreover, even reverses the trend, resulting in an effective red shift upon increasing the coupling strength.\\
We analysed the azimuthal angle dependent maximum energy position of the PL-peak and observed a sine like behaviour between 1.853\,eV and 1.857\,eV (orange curve in Figure S7). In contrast, the absorption remains unchanged at a mean value of 1.931\,eV (blue curve ibid.). Subtracting both values and thus obtaining the effective plexcitonic Stokes-shift as function of azimuthal angle, the sinusoidal fit reveals an energy amplitude of $\approx3$\,meV and a $180^\circ$ symmetry as expected from the absorption. This Stokes-shift includes all previously discussed effects, namely the combined shift of $E_{Stokes,Plex}$, energetic splitting $\Omega_{1,2}$ and the energy shift originating from the relaxation paths displayed in Figure \ref{Fig.: Result Emission Plexciton Dispersion}a).\\
The PL-dispersion simultaneously leads to a point where the FWHM of the emission peak becomes narrower, see Figure \ref{Fig.: Result Emission Plexciton Dispersion}c). The local minimum of the FWHM at $k=10.25 \dfrac{1}{\mu \text{m}}$ coincides with the maximum of the PL intensity and hence, marks the optimal balance between the coherently coupled excitonic part of the PBI-aggregate and the plasmonic part represented by the SPP, i.e. between the delocalizing character of the plasmon and the energetic relaxation properties of the exciton \cite{wei2020overcoming, PhysRevLett.108.066401}. Remarkably, this $k$-value is not coinciding with the energetic maximum of the PL dispersion but is shifted by $\Delta k=0.24 \dfrac{1}{\mu \text{m}}$. We assume this in favor of the coherently coupled J-aggregates to be in a more delocalized plexcitonic state, which is mainly mediated by its plasmonic part and yields the observed $k$-value shift towards a plexciton state with higher plasmonic weighting. Analyzing the dependence of the photophysical behavior on the light-matter-character we identified a sweet spot at $77\%$ plasmonic and $23\%$ excitonic contribution where the occurring delocalization suggests a more coherent emission accompanied by a decrease in FWHM and increase of the PL-intensity. This results in the first PL-intensity maximum at low $k$ in Figure \ref{Fig.: Result Emission Plexciton Dispersion}d) \cite{wei2020overcoming, PhysRevLett.108.066401}. While the FWHM approaches a fixed value of 140\,meV, which resembles the width of 155\,meV for the neat molecular exciton in thin films (see Figure \ref{Fig.: Experimental and Methods}c)), the subsequent local minimum accompanied by a second local maximum in the PL-intensity in Figure \ref{Fig.: Result Emission Plexciton Dispersion}d) is rather unexpected.\\
To further elucidate the PL-intensity trend in Figure \ref{Fig.: Result Emission Plexciton Dispersion}d), we modelled the photoluminescence of the plexcitonic state for the same four different cases as we did for the absorption data (see Figure S9). Further, we only considered those cases where the anti-crossing is, at least, 8\,meV in order to take only the plexciton splittings into account which we also see in our experimental data. It is noticeable that PL-plexciton-peaks will arise for each $k$-value where an anti-crossing can be observed. We can distinguish between two main contributions to the emission signal: First, an isotropic fraction being independent of the azimuthal angle and, second, a major contribution showing a pronounced dependency on the azimuth at the anti-crossing. In our data we also observe multiple peaks as shown in Figure \ref{Fig.: Result Emission Plexciton Dispersion}d). While the second peak maximum at $k=11.2 \dfrac{1}{\mu \text{m}}$ is matching to the weak coupling modulation above the main anti-crossing, the first peak shows quite a significant shift of approximately $\Delta k=0.44 \dfrac{1}{\mu \text{m}}$. We attribute this $k$-value deviation to the angle dependent out-coupling efficiency of the emitting dipole transition\cite{frischeisen2010determination} and to aforementioned emission of the coherently coupled J-aggregate contribution to the plexciton, ultimately boosting and shifting the PL-intensity for lower $k$-values. Quite noticeable is the simultaneous decrease in intensity of the anisotropic (first peak) and isotropic (second peak) contributions ($45\%$ intensity ratio of strong-to-weak coupling case), which we could also reproduce in our calculations, but with different absolute values. This hints at a strong relation between the anti-crossing at higher energy and that at 1.93\,eV, which is constituted by the relaxation from the higher into the lower plexciton state, followed by the radiative decay into the ground state leading to the second PL-peak.\\
\section{Power Dependent Studies}
\begin{figure}
\includegraphics[]{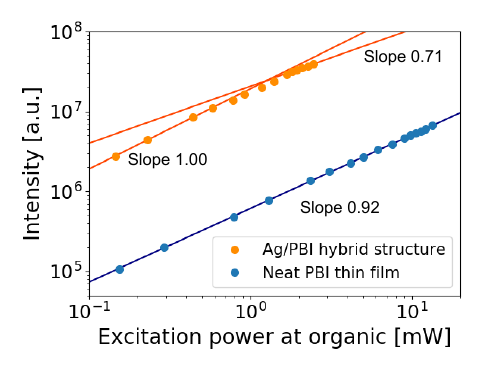}
\centering
\caption{PL-intensity of the Ag/PBI plexciton sample (orange dots) and a neat PBI reference sample (blue dots) on glass as function of effective excitation power (corrected for the silver layer absorption in case of the former). This power-series was measured for the strong coupling case at $\varphi=0^\circ$ at an polar angle at $\theta=46^\circ$. The linear fits of slope 1 and 0.92 indicate the single particle character of excitation and emission in the two different samples. Besides the much higher quantum efficiency for the metal-organic hybrid sample, which we attribute to the suppression of non-radiative decay channels, the decrease in slope at excitation powers above 20\,mW is considered indicative for the onset of plexciton-plexciton annihilation which, according to the low power threshold, pinpoints at a strongly enhanced plexciton diffusion caused by the propagating character of the plasmonic contribution \cite{cook2010singlet, dostal2018direct, suna1970kinematics, marciniak2011one}.}
\label{Fig.: Power PL Study}
\end{figure}

Based on our previous findings on the static optical properties, we extent our analyses in the following to the plexciton transport and its dependence on the light-matter coupling. For this purpose, we carried-out PL measurements at different excitation powers, at fixed polar angle $\theta=46^\circ$ and explicitly accounting for the strong coupling case at $\varphi=0^\circ$, which corresponds to the first PL-intensity maximum in Figure \ref{Fig.: Result Emission Plexciton Dispersion}d). The results are shown in a double logarithmic plot in Figure \ref{Fig.: Power PL Study} together with a neat molecular PBI film, acting as reference and measured under the same conditions as our metal-organic hybrid structure. To compare the intensities, we had to correct the excitation and emission of the plexciton sample for the absorption of the silver layer ($A\approx0.9$). At first glance, a comparison of both power-dependent intensity curves immediately reveals a strongly enhanced PL by more than one order of magnitude in case of the metal-organic hybrid structure. Moreover, both intensity curves can be linearly fitted over an extended power range with slopes of 1 and 0.92, respectively, confirming the excitation and emission process to be of single particle character. As such an emission enhancement of similar amplitude has been reported already for many other combinations of (organic) semiconductors and metals in close proximity \cite{kolb2017hybrid, russell2012large, ming2012plasmon} we anticipate the same fundamental mechanism at work, viz. the increase in the radiative transition rate by the strongly enhanced final state density caused by the metal and its collective states.\\
Furthermore, at highest excitation powers we had to account for the intensity reduction by photobleaching in the neat PBI reference sample on glass. In contrast, for the silver-PBI plexciton sample no such bleaching was visible in the applied excitation power range, hinting at a strong delocalization and/or shorter lifetime of the excited state. For a pure molecular aggregate, a reduction in radiative lifetime would lead to an increase in PL intensity due to the suppression of long-lived non-radiative decay paths and the faster re-excitation feasible. However, since the power-dependent PL intensity of the metal-organic hybrid structure does not show such an additional increase but, on contrary, even starts to decrease in slope to 0.71 (see Figure \ref{Fig.: Power PL Study}), we must refer the reduction in lifetime to a different origin. In particular, the reduction in slope hints to an onset of efficient plexciton-plexciton (plex-plex) annihilation in the plexciton sample, where we would anticipate a slope of 0.5 for the pure two-particle process \cite{cook2010singlet, dostal2018direct, suna1970kinematics, marciniak2011one}.\\
Based on this interpretation of the power-dependent intensity data, in the next step we will estimate the underlying plexciton diffusion length and compare it with data known for other long-range ordered molecular structures. For this purpose, we utilize as starting point the equation for the time-dependent local exciton/plexciton density $n(x,t)$, based on Fick’s law with additional generation and decay terms
\begin{equation}
\frac{\delta n}{\delta t}= D \nabla^2 n- \frac{n}{\tau} - \gamma n^2 + G
\end{equation}
which, under steady-state conditions (i.e. at constant exciton generation $G$ and $\dfrac{\delta n}{\delta t}= \nabla^2 n=0$) transforms to
\begin{equation}
n=\frac{1}{2 \gamma \tau}\left(\sqrt{1+4 \gamma \tau^2 G}-1\right).
\end{equation}
Here, $\gamma$ is the annihilation rate and, $\tau$ the lifetime of the excitonic/plexcitonic state. Since the precise exciton lifetime in our PBI-based J-aggregate under study is not known we will approximate this value by literature data for similar PBI aggregates, taking also into account their reduction by the enhanced radiative decay rate due to the enhanced final state density of the metal film close by (see also discussion above).\\
Table \ref{tbl: Plex-Plex Annihiliation Fit Values} shows a comparison of the mean literature value\cite{dostal2018direct, lin2010collective} and our parameters deduced by fitting the power dependent intensity curves in Figure \ref{Fig.: Power PL Study} with our diffusion model. Following the enhancement of the PL intensity by more than one order of magnitude we assume a drastically shortened lifetime of about 75\,ps for the excited state in the silver-PBI hybrid stack. This shortening can be further rationalized by the fact that the plexcitonic state is a combination of plasmonic excitation (typical lifetime in the ps-range\cite{mahan2018lifetime}) and molecular excitation (lifetime of 730\,ps for similar PBI derivatives). Utilizing this value of 75\,ps and considering $G_{crit}=\dfrac{1}{\gamma \tau^2}$, which is the threshold generation rate for the annihilation process, and $R_a= 4.02\,\text{\AA}$ which is the annihilation radius of two plexcitons (commonly approximated by the lattice distance between two molecules), the diffusion constant $D=\dfrac{\gamma}{4\pi R_a}$ and thus the 1D diffusion length $L_d= \sqrt{2 \tau D}$ in the highly elongated PBI strands can be calculated.\\
\begin{table}
  \caption{Photophysical properties of neat PBI molecules (lifetime taken from literature\cite{dostal2018direct, lin2010collective}) and in Ag/PBI hybrid layers.}
  \label{tbl: Plex-Plex Annihiliation Fit Values}
  \begin{tabular}{|c|c|c|}
    \hline
      & Literature PBI\cite{dostal2018direct, lin2010collective} & Plexciton \\
    \hline
    \hline
    $\tau$ & 730\,ps & 75\,ps \\
    \hline
    $\gamma$ & $2.6\times10^{-14}$\,m$^3$/s & $9.9\times10^{-14}$\,m$^3$/s \\
    \hline
    $D$ & 0.03\,cm$^2$/s & 2\,cm$^2$/s \\
    \hline
    $L_d$ & 70\,nm & 172\,nm \\
    \hline
  \end{tabular}
\end{table}The lifetime assumption of the hybrid state is feasible\cite{gomez2013picosecond} and is supported by our measurements since the total intensity (after correcting the absorbing silver layer) is significantly higher. The substantially higher diffusion constant for the plexcitonic excitation compared to diffusion constants for neat molecular excitons, either PBI or many other crystalline organic semiconductors\cite{kurrle2008exciton, mikhnenko2012exciton, lunt2010relationship, mikhnenko2015exciton} together with the larger diffusion length can be rationalized by the plasmonic character of the hybrid state, which by its propagating nature and its mean free path of several micrometer\cite{kolomenski2009propagation, lamprecht2001surface, kolomenski2009propagation} supports the transport of the excitation and its correlated energy. Based on this data, the onset of plexciton-plexciton annihilation already at smaller excitation powers can be ascribed to the large diffusion length in combination with an enhanced delocalization of the plexcitonic state in contrast to the strongly localized Frenkel or charge transfer state of the pure PBI J-aggregates. With regard to a technological perspective, the superior transport of hybrid states in light-matter coupled material systems\cite{orgiu2015conductivity} offers new possibilities not only to control the spatially anisotropic optical excitations but also to steer their energy in certain directions. This could be of relevance for current concepts of photovoltaic devices or photodectors as well as for future applications in e.g. highly integrated photonic circuitries.\\
\section{Conclusion}
To summarize, we have demonstrated in our work how by carefully selecting suited molecular semiconductors with spatially anisotropic packing and discrete optical excitations new static as well as dynamic optically characteristics can emerge upon coupling to the dispersive excitation of plasmonic structures. The new optical states are of mixed excitonic and plasmon-polaritonic nature which leads to lifting of the degenerated states via anti-crossing in the absorption at distinct energy and momentum. The angular dependence of the coupling allows for a variation of the splitting between 0 and 28\,meV for the given Ag/PBI hybrid layers and, thus, provides a platform for studying secondary effects by the coupling, e.g. upon forming crystalline organic donor-acceptor junctions on top of the layer. Moreover, our combined analysis of the angular and wavevector dependent absorption and photoluminescence show clear evidence for coherent processes of energy exchange between the excitonic and the plasmonic part of the plexciton. This, on one hand effects the fundamental optical properties of the hybrid state, like its Stokes shift which depends on the respective coupling strength and, thus, allows for conceptualizing the presented excitation and relaxation scheme. Finally, also for the plexciton transport, their mixed nature and the inherent light-matter coupling lead to a strongly enhanced, directional diffusion outperforming those of neat molecular excitons by more than an order of magnitude. In combination, these specific hybrid state properties are considered extremely promising for a fundamental understanding of the photophysics in such coupled metal-organic structures and their further optimization towards current and future opto-electronic applications.

%%%%%%%%%%%%%%%%%%%%%%%%%%%%%%%%%%%%%%%%%%%%%%%%%%%%%%%%%%%%%%%%%%%%%
%% The "Acknowledgement" section can be given in all manuscript
%% classes.  This should be given within the "acknowledgement"
%% environment, which will make the correct section or running title.
%%%%%%%%%%%%%%%%%%%%%%%%%%%%%%%%%%%%%%%%%%%%%%%%%%%%%%%%%%%%%%%%%%%%%
\begin{acknowledgement}
All authors acknowledge financial support by the Bavarian State Ministry for Science and the Arts within the collaborative research network ``Solar Technologies go Hybrid'' (SolTech). L.N.P. acknowledges a Kekulé fellowship by the Fonds der chemischen Industrie
\end{acknowledgement}

%%%%%%%%%%%%%%%%%%%%%%%%%%%%%%%%%%%%%%%%%%%%%%%%%%%%%%%%%%%%%%%%%%%%%
%% The same is true for Supporting Information, which should use the
%% suppinfo environment.
%% A listing of the contents of each file supplied as Supporting Information should be included. For instructions on what should be included in the Supporting Information as well as how to prepare this material for publications, refer to the journal's Instructions for Authors.
%%%%%%%%%%%%%%%%%%%%%%%%%%%%%%%%%%%%%%%%%%%%%%%%%%%%%%%%%%%%%%%%%%%%%

%%%%%%%%%%%%%%%%%%%%%%%%%%%%%%%%%%%%%%%%%%%%%%%%%%%%%%%%%%%%%%%%%%%%%
%% The appropriate \bibliography command should be placed here.
%% Notice that the class file automatically sets \bibliographystyle
%% and also names the section correctly.
%%%%%%%%%%%%%%%%%%%%%%%%%%%%%%%%%%%%%%%%%%%%%%%%%%%%%%%%%%%%%%%%%%%%%
\bibliography{PaperBibliography}

\end{document}

\renewcommand{\thefigure}{S\arabic{figure}}
\renewcommand{\thetable}{S\arabic{table}}
\renewcommand{\theequation}{S\arabic{equation}}
\part*{Supporting Information}
\section{Materials}
Tetra-bay phenoxy-dendronized PBI \textbf{1} was synthesized and the aligned liquid-crystalline layer of J-aggregates was prepared according to our previous described methodology.\cite{herbst2018self,kaiser2009fluorescent}
\section{Off-Centered Spin Coating}
\begin{figure}[H]
\includegraphics[width=0.99\textwidth]{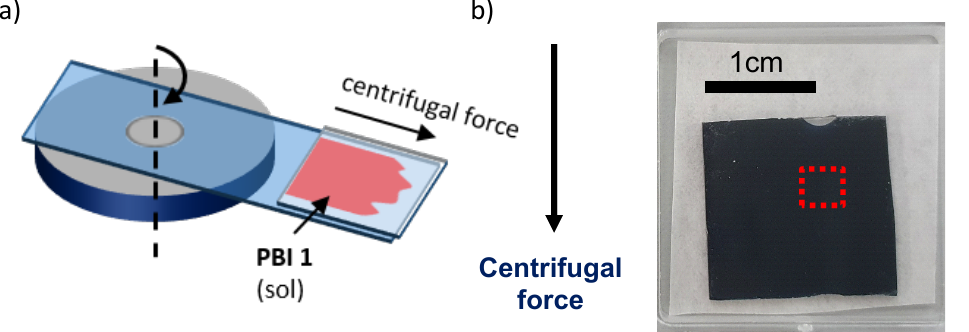}
\caption{a) Schematic illustration of the off-centered spin-coating method for the preparation of aligned J-aggregates of PBI \textbf{1}.\cite{kim2022wavelength, yuan2014ultra} b) Real image of the measured sample after spin-coating. The red area indicates the measuring spot of the polarization-dependent fluorescence microscopy. \label{S_fig:Off Centered Spin Coating} }
\end{figure}
For the anisotropic sample preparation we used the off-center spin-coating method\cite{kim2022wavelength, yuan2014ultra}  as displayed in figure \ref{S_fig:Off Centered Spin Coating}. The sample is placed with its center approximately 3\,cm away, in this way the centrifugal force ensures the alignment of the 1D helical strands of the PBI-Aggregates along the centrifugal force vector. The previously fabricated 45\,nm Ag/Glass substrates via thermal evaporation were cleaned in acetone and isopropyl alcohol within a ultrasonic bath for 10 minutes, respectively. After that the thin PBI layer is spin-cast from a 90\,$^\circ$C hot chlorobenzene solution with a concentration of 4\,$\frac{\textit{mg}}{\textit{mL}}$. The acceleration speed was kept at 500\,rpm while the final speed and total time was 2000\,rpm and 30\,seconds, respectively. The used sample for the measurements in the main text is shown in figure \ref{S_fig:Off Centered Spin Coating}b) with the indicated centrifugal force direction. A 15-20\,nm thick and well-aligned PBI film is formed on top of the 45\,nm thick silver layer. The red dashed marked area indicates the measured spot for the polarization-dependent fluorescence microscope image (figure 1b in the main text). For better comparison the orientation was kept the same for both images. As a measure for the molecular orientation within the organic thin film the dichroic ratio
\begin{align}
D_\lambda =\dfrac{A_{\lambda , 0^\circ}}{A_{\lambda , 90^\circ}}
\end{align}
can be calculated, with $A_{\lambda , 0^\circ}$ and $A_{\lambda , 90^\circ}$ being the absorption at a certain wavelength $\lambda$ for parallel and perpendicular orientation with regard to the alignment, respectively.

\section{Surface Plasmon Dispersion of Silver}
\begin{figure}[H]
\includegraphics[width=0.99\textwidth]{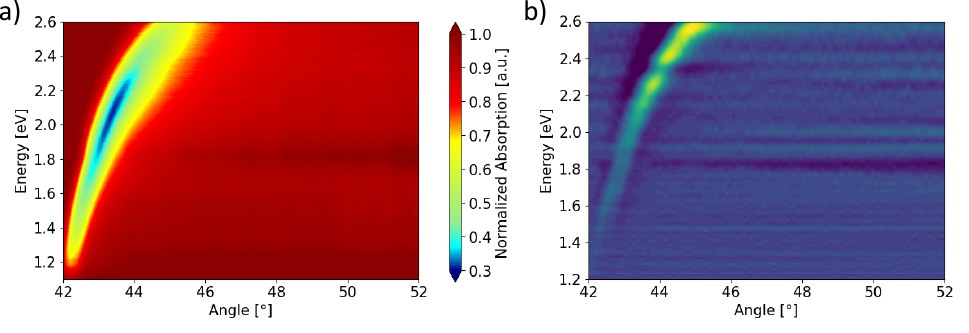}
\caption{Reference surface plasmon disperison of a 45\,nm thick silver thin film on glass. a) Normalized absorption spectrum. b) Second derivative of the normalized absorption data. \label{S_fig:Silver Reference Spectra} }
\end{figure}

In figure \ref{S_fig:Silver Reference Spectra} the uncoupled surface plasmon of a 45\,nm thin silver film is shown. We evaporated the thin films via thermal evaporation as it is described in the main text in order to get a reference spectra for comparison with the plexciton spectra. As can be seen in the second derivative, contrary to the normalized absorption spectra in figure \ref{S_fig:Silver Reference Spectra}a), a small feature at 2.34\,eV is already visible without the PBI film on top. We believe that the real signal in the main text consists of both features: The artificial signal which is visible in figure \ref{S_fig:Silver Reference Spectra}b) and a superimposed anti-crossing feature.

\section{Optical and Structural Properties of PBI J-Aggregate}

\subsection{Absorption Characteristics}
\begin{figure}[H]
\includegraphics[width=0.9\textwidth]{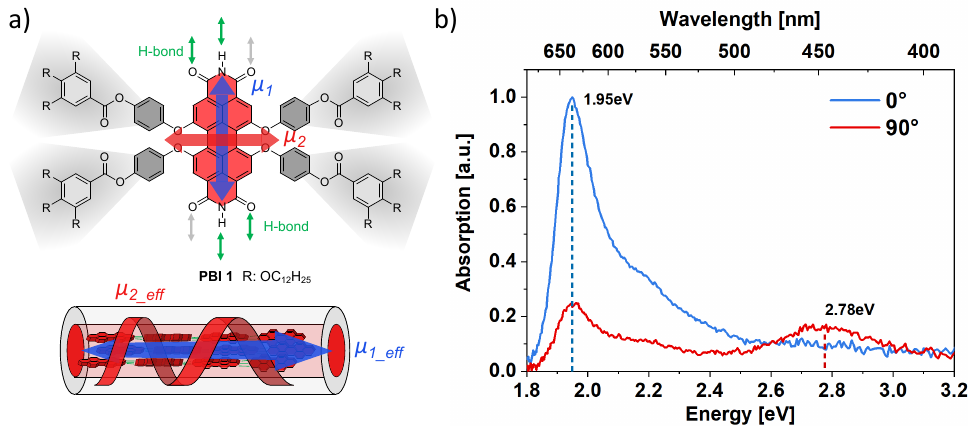}
\caption{a) Molecular transition dipole moment orientation of the PBI \textbf{1} (top) as well as for the J-Aggregates(bottom) within the triple-stranded helix for the high energy transition $\mu_2$ (red) and the lower energetic exciton transition $\mu_1$ (blue). The corresponding effective dipole transition orientation is illustrated below. b) Polarization-dependent UV-Vis absorption spectra for 0\,$^{\circ}$ and 90\,$^{\circ}$ light polarisation of an aligned thin film of J-aggregates with respect to the alignment. \label{S_fig:Angle Dependend PBI Absorption} }
\end{figure}

The polarization-dependend UV-Vis absorption spectra at normal incidence for the J-Aggregate helical structure is shown in figure \ref{S_fig:Angle Dependend PBI Absorption}b). As can be seen the optimal excitation of the S\textsubscript{0}-S\textsubscript{1} as well as S\textsubscript{0}-S\textsubscript{2} transitions at 1.95\,eV and 2.78\,eV, respectively, differ by 90$^\circ$ originating from the perpendicular alignment of respective transition dipole moments within the PBI chromophore in figure \ref{S_fig:Angle Dependend PBI Absorption}a) (indicated in red and blue). While the lower energetic transition dipole $\mu_1$ lines up with the long molecular N,N´ axis (blue arrow) the high energetic transition $\mu_2$ is placed alongside the short molecular axis (red arrow).
The resulting effective transition dipole moment orientation $\mu_{1\_eff}$ and $\mu_{2\_eff}$ therefore exhibit a distinct orientation in the helical triple-strand depicted in figure \ref{S_fig:Angle Dependend PBI Absorption}a). While the lower excited state is oriented along the triple-stranded long columnar axis the energetic higher transition rotates around the $\mu_{1\_eff}$ axis and subsequently grantees a partial projection of $\mu_{2\_eff}$ on the x- y- plane enabling our photoluminescence experiment.

\subsection{Hopfield Coeficient of the Plexcitonic State}
\begin{figure}[H]
\includegraphics[width=0.6\textwidth]{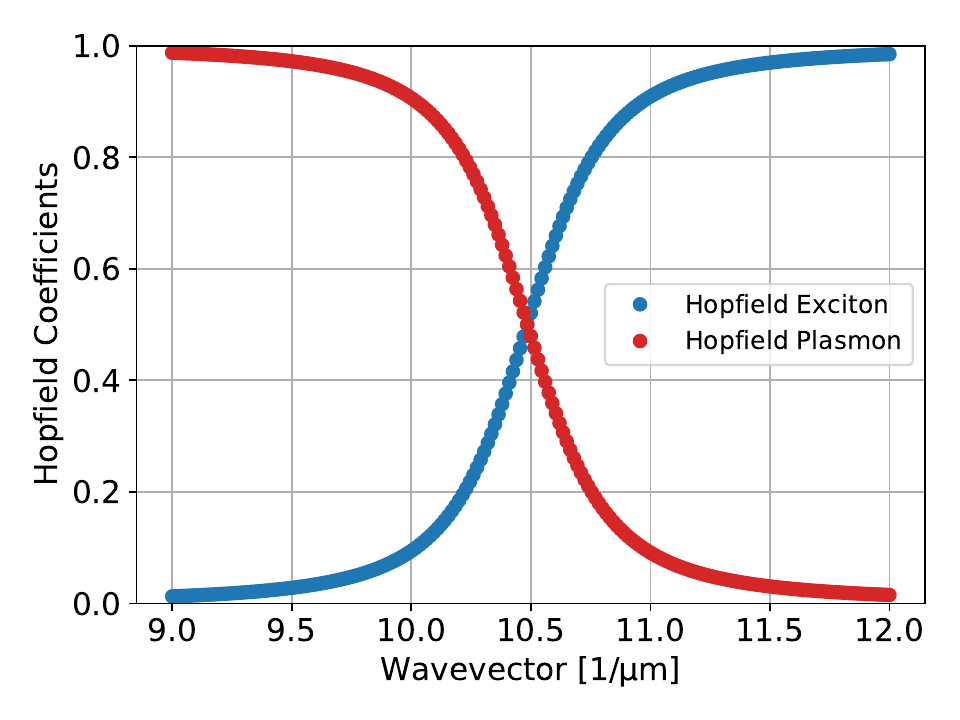}
\caption{Calculated Hopfield Coefficients of the plexcitonic state at 1.93\,eV, indicating the plascmonic and the excitonic part of the coupled plexciton. \label{S_fig:Hopfield_Coefficients} }
\end{figure}

In order to quantify the anti-crossing we calculated the Hopfield coefficients. The result for the strongest coupling strength of 28\,meV at an azimuthal angle of 90\,$^\circ$ at 1.93\,eV is shown in figure \ref{S_fig:Hopfield_Coefficients}. The wave vector for a 50/50 excitonic/plasmonic ration can be determined to 10.49\,$\frac{1}{\mu\textit{m}}$.

\subsection{Geometry-Optimized Model of the Helical Structure}
\begin{figure}[H]
\includegraphics[width=0.9\textwidth]{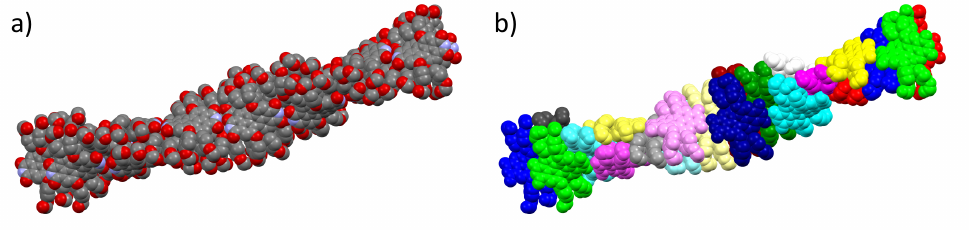}
\caption{Optimized self-assembled structure of the helical triple-stranded PBI J-aggregate without the large solubilizing side chains in the periphery. a) Colour-coded by elements (C: grey; O: red; N: blue) b) Colour-coded by molecules. \label{S_fig:Theoretical Helix}}
\end{figure}
The optimized self-assembled structures of the triple-stranded J-aggregate structure of a simplified  PBI \textbf{1} without the aliphatic chains based on wide and middle angle X-ray scattering (WAXS/MAXS) measurements was carried out via BIOVIA Materials Studio 2017 R2. The helical structure was assembled as previously described and geometry optimised via the module Forcite with the COMPASS II forcefield using the Ewald summation method. This was performed until the optimisation converged with high negative Non-bonding intermolecular interaction energies (van der Waals and electrostatic interactions).\cite{herbst2018self}
\begin{table}[H]
\begin{tabular}{|c|c|c|c|c|c|c|}
\hline 
 &  &  & Lattice Number &  &  & \tabularnewline
 &  &  & of mol. per &  &  & \tabularnewline
 T [$^\circ$C] & Phase & a [\AA] & col. Stratum & $\pi-\pi$ [\AA] & Helix & Pitch [\AA]\tabularnewline
\hline 
\hline 
 180 & Col\textsubscript{h} & 40.2 & 3 & 4.1 & $7_1$ & $7\cdot 14.1=98.7$\tabularnewline
\hline
\end{tabular}
\caption{Structural parameters of the generated geometry-optimized models
of the helical structure of PBI \textbf{1} J-aggregates based on previously reported WAXS/MAXS measurements and modelled with the program Accelrys Materials Studio 4.4.\cite{herbst2018self} \label{S_Tab:Theoretical Values} }
\end{table}

\section{Photoluminescence Spectra of the Plexcitonic State}
\begin{figure}[H]
\includegraphics[width=0.9\textwidth]{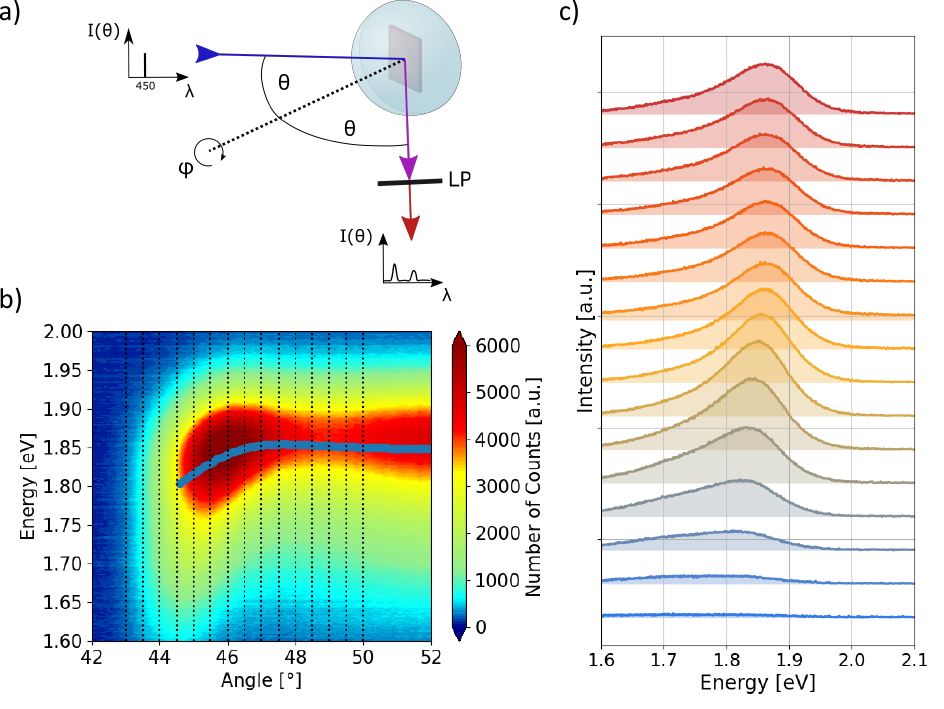}
\caption{a) Set-Up for measuring the plexciton photoluminesence in dependence of the azimuthal angle $\varphi$. b) Plexciton photoluminescence for the strongest coupling strength at an azimuthal angle of 0\,$^\circ$ of an aligned sample of PBI \textbf{1} J-aggregates on an silver surface. The dotted vertical lines indicate the discplayed spectra in c). The blue dots in the graph are the maximum position of the corresponding fit. c) Waterfall plot for the indicated spectra in b).    \label{S_fig:PL Plexciton}}
\end{figure}

\begin{figure}[H]
\includegraphics[width=0.4\textwidth]{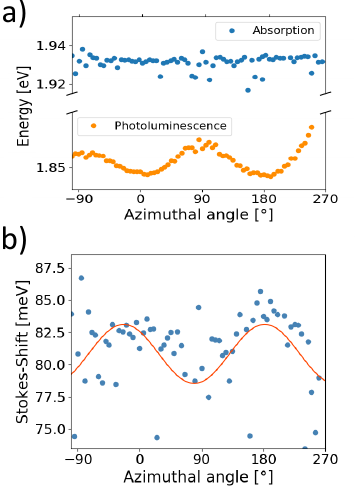}
\caption{a) Azimuthal angle dependent change of the energetic position for the absorption (top) and PL (bottom) plexciton dispersions. While the energetic position of the absorption is extracted from the fit of the respective plexciton branch, the energy position of the PL is taken from the maxima of the curves, i.e. in figure 6b). b) Angular dependent Stokes shift calculated by the energetic values in a).\label{S_fig:Stokes-Shift}}
\end{figure}

The angle-dependent PL-measurements setup is displayed in figure \ref{S_fig:PL Plexciton}a). By exchanging the white light source with a 450\,nm laser and additionally placing a longpass filter in the beam path we were able to measure the PL-response of the plexciton. Figure \ref{S_fig:PL Plexciton}b) show such a PL-heatmap for the strongest coupling at an azimuthal angle of 0\,$^\circ$. The dotted black lines indicate  used spectra for the waterfall plot in \ref{S_fig:PL Plexciton}c).
For further evaluation we fitted each spectra with a Lorentzian profile and extracted the characteristic peak position (plotted in the main text in figure 6b)), full with at half maximum (FWHM) (figure 6c) in the main text) and the peak area of the fit (figure 6d) in the main text) as a measure for the intensity. The energetic position for each fit is plotted in figure \ref{S_fig:PL Plexciton}b) in blue as well.
%\begin{figure}[H]
%\includegraphics[width=0.6\textwidth]{S_8.pdf}
%\caption{Fitting results of the strongest coupling ($\vec{k} \parallel \mu $, green), weakest coupling ($\vec{k} \perp \mu$, blue) and an intermediated step (orange). a) Angle-dependent peak maxima, b)FWHM of the fitted peaks and c) shows the peak area (amplitude) of the corresponding fitted peaks.\label{S_fig:PL Plexciton_eval}}
%\end{figure}
%A more detailed analysis of the fitted data for all azimuthal angles can be seen in figure \ref{S_fig:PL Plexciton_eval} not only for the strongest coupling at an azimuthal angle of 0\,$^\circ$ but also the weakest coupling at $\varphi=90\,^\circ$ and an intermediated step at $\varphi=45\,^\circ$. The graph shows the values like FWHM and the peak area as a measure for the intensity. The minima at the FWHM hints to coherent effects and coincides with the PL-maxima shown below. Both features are shifted towards higher $k$ values as discussed in the main text.\\
Additionally we plotted the azimuthal angle dependent maximal energetic position of the PL-peaks for the absorption and emission plexciton dispersion in figure \ref{S_fig:Stokes-Shift}a), in order to calculate the effective Stokes shift of our plexcitonic states. The resulting Stokes shift can be seen in figure \ref{S_fig:Stokes-Shift}b) together with a cosine model fit, indicated by the orange solid line.

\section{Theoretical Methods}
\subsection*{Exciton Hamiltonian}
In this section a theoretical framework for the description of excitons is described. Out of convenience operators will not be marked with a hat in this section. The Hamiltonian of an exciton consisting of $N$ molecules can be written as follows

\begin{equation}
    H_{\mathrm{exc}} = \sum_{i=1}^N H_i + \sum_{i=1}^N \sum_{j>i} V_{ij}
\end{equation}

with the single molecule Hamiltonians $H_i$ and the pairwise couplings between the molecules $V_{ij}$. These interactions are mediated by Coulomb forces between electrons and nuclei of different monomers and thus are given by (in atomic units)

\begin{equation}
    V_{ij} = \sum_{a\in i}\sum_{b\in j} \frac{1}{|\Vec{r}_a-\Vec{r}_b|} - \sum_{A\in i}\sum_{b\in j} \frac{Z_A}{|\Vec{R}_A - \Vec{r}_b|} - \sum_{a\in i}\sum_{B\in j} \frac{Z_B}{|\Vec{r}_a - \Vec{R}_B|} + \sum_{A\in i}\sum_{B\in j} \frac{Z_AZ_B}{|\Vec{R}_A - \Vec{R}_B|},
\end{equation}

where the individual sums go over all electrons (nuclei) of one monomer, $\Vec{r}$ ($\Vec{R}$) is the electron (nuclear) coordinate and $Z$ is the charge of a nucleus. A basis of the Hilbert space of the exciton Hamiltonian is constructed by taking direct products of the eigenstates of the single molecule Hamiltonians $\left\{|\Psi^i_\alpha\rangle,\ \alpha = 0,\ \dots,\ M_i\right\}$ with the number of eigenstates of the i-th molecule $M_i$

\begin{equation}
    \left\{|\phi_{\alpha, \dots, \gamma, \dots, \zeta}\rangle\right\} = \left\{|\Psi^1_\alpha\rangle \otimes \cdots \otimes |\Psi^i_\gamma\rangle \otimes \cdots \otimes |\Psi^N_\zeta\rangle\right\}.
\end{equation}

This basis consists of $\Pi_{i=1}^N M_i$ basis states and is therefore computationally difficult to handle with a large number of monomers and eigenstates per monomer. Thus, in the following only basis states which involve a single excited monomer are considered, i.e.

\begin{equation}
    \left\{|\phi_{0, \dots, \gamma, \dots, 0}^c\rangle\right\} = \left\{|\Psi^1_0\rangle \otimes \cdots \otimes |\Psi^c_\gamma\rangle \otimes \cdots, \otimes |\Psi^N_0\rangle\right\}
\end{equation}

where the superscript $c$ denotes the excited molecule. Furthermore, depending on the system under consideration and the scientific question, the number of eigenstates per monomer has to be chosen mindfully. To further simplify the description of the interaction, only matrix elements of basis states, which carry the excitation on different monomers, are considered. These matrix elements are given by

\begin{equation}
    \langle\phi_{0, \dots, \delta, \dots, 0}^j|V_{ij}|\phi_{0, \dots, \gamma, \dots, 0}^i\rangle = \int \mathrm{d}\Vec{r}_a \mathrm{d}\Vec{r}_b \frac{\rho^i_{0\gamma}(\Vec{r}_a)\rho^j_{0\delta}(\Vec{r}_b)}{|\Vec{r}_a - \Vec{r}_b|}.
\end{equation}

This expression is derived by using the orthogonality and antisymmetry of the single molecule eigenstates, and the transition densities between ground and excited states were introduced

\begin{equation}
    \rho_{0\gamma}^i (\Vec{r}_1) = \int \cdots \int \mathrm{d}\sigma_1 \mathrm{d}\Vec{x}_2 \dots \mathrm{d}\Vec{x}_{N_i} \Psi_0^{i*}(\Vec{x})\Psi_\gamma^i(\Vec{x}),
\end{equation}

where $\Vec{x}$ are all combined spin and spatial coordinates $\left\{\Vec{x}_n\right\}$ of all electrons of the monomer $i$, $\sigma_1$ is the spin coordinate of the first electron, $N_i$ is the number of electrons and $\Psi_0^i(\Vec{x})$ ($\Psi_\gamma^i(\Vec{x})$) is the ground (excited) state wave function of monomer $i$. Since the calculation of the matrix elements of equation (5) is computationally demanding for large molecules, the transition charge approximation is employed \cite{madjet2006intermolecular}. Therefore the exciton coupling elements are written in the form

\begin{equation}
    \langle\phi_{0, \dots, \delta, \dots, 0}^j|V_{ij}|\phi_{0, \dots, \gamma, \dots, 0}^i\rangle = \int \mathrm{d}\Vec{r}_a \rho_{0\gamma}^i(\Vec{r}_a)\phi_{0\delta}^j(\Vec{r}_a),
\end{equation}

here the electrostatic potential $\phi_{0\delta}^j(\Vec{r}_a)$ of the corresponding transition density is introduced, which is given by

\begin{equation}
    \phi_{0\delta}^j(\Vec{r}_a) = \int\mathrm{d}\Vec{r}_b \frac{\rho^j_{0\delta}(\Vec{r}_b)}{|\Vec{r}_a-\Vec{r}_b|}.
\end{equation}

The key idea is now to approximate the electrostatic potential by atom centred transition charges $q_{0\delta}^{i(B)}$

\begin{equation}
    \phi_{0\delta}^j(\Vec{r}_a) \approx \sum_{B\in j} \frac{q_{0\delta}^{j(B)}}{|\Vec{r}_a-\Vec{R}_B|}.
\end{equation}

Applying this approximation to both transition densities in equation (7), leads to

\begin{equation}
     \langle\phi_{0, \dots, \delta, \dots, 0}^j|V_{ij}|\phi_{0, \dots, \gamma, \dots, 0}^i\rangle = \sum_{A\in i} \sum_{B\in j} \frac{q_{0\delta}^{j(B)}q_{0\gamma}^{i(A)}}{|\Vec{R}_A - \Vec{R}_B|}.
\end{equation}

The form of the exciton coupling in terms of the atomic transition charges is especially simple because it only involves Coulomb interactions between point charges, which are easily calculable. There are multiple ways to obtain atomic transition charges from quantum chemical methods, which will not be discussed here \cite{renger2012theory, hirshfeld1977bonded}. 

\subsection*{Plexciton Coupling}

In the following a theoretical framework for the description of the light matter interaction of the excitonic structure with surface plasmon polariton modes will be derived. Since the interaction of the surface plasmon polariton with the molecular exciton in the dipole approximation only depends on the electric field, we will focus on the properties of the latter in the following. The component of the electric field perpendicular to the interface shows an exponential decay into both materials and usually decays faster in the metal than in the dielectric material. Furthermore, this field component is enhanced compared to the component parallel to the surface. Due to this strong confinement of the surface plasmons polariton fields, these have to be quantized. This can be done by the standard procedure of quantum mechanics \cite{gonzalez2013theory}. The Schr\"{o}dinger operator of the quantized electric field strength in the dielectric medium of one mode with the in-plane wave vector $\Vec{k}$ is given by

\begin{gather}
    \hat{\Vec{E}}_{\mathrm{SPP}} = \sqrt{\frac{\hbar\omega(\Vec{k})}{2\epsilon_0 A}} \Vec{\epsilon}_{\Vec{k}} \mathrm{e}^{-k_z z} \left(\hat{a}_{\Vec{k}}\mathrm{e}^{\mathrm{i}\Vec{k}\Vec{x}} + \hat{a}_{\Vec{k}}^\dag \mathrm{e}^{-\mathrm{i}\Vec{k}\Vec{x}}\right),
\end{gather}

with the quantized area $A$, the vertical component of the wave vector $k_z$ and the distance from the interface $z$. The term $\mathrm{e}^{-k_z z}$ describes the exponential decay of the fields into the materials. The in-plane wave vector $\Vec{k}$ and the in-plane position $\Vec{x}$ are two dimensional vectors in the plane parallel to the interface. The non-normalized polarization vector $\Vec{\epsilon}_{\Vec{k}}$ is given by

\begin{equation}
    \Vec{\epsilon}_{\Vec{k}} = \Vec{u}_{\Vec{k}} + \mathrm{i}\frac{|\Vec{k}|}{k_z}\Vec{u}_z,
\end{equation}

where $\Vec{u}_z$ is the unit vector in the z-direction and $\Vec{u}_{\Vec{k}}$ is a unit vector in an in-plane direction. The factor $\frac{|\Vec{k}|}{k_z}$ usually leads to the enhancement of the electric field perpendicular to the surface, in the following it will be denoted as the enhancement constant $d$. $\omega(\Vec{k})$ is the plasmon dispersion relation, whose inverse is given by \cite{kretschmann1968radiative}

\begin{equation}
    |\Vec{k}| = \frac{\omega}{c}\sqrt{\frac{\epsilon_1\epsilon_2}{\epsilon_1+\epsilon_2}},
\end{equation}

with the dielectric function of the metal (dielectric) $\epsilon_2$ ($\epsilon_1$). The dielectric function of the dielectric is assumed to be constant over the given interval and the dielectric function of the metal in the free electron model of an electron gas is given by \cite{ritchie1957plasma}

\begin{equation}
    \epsilon_2\left(\omega\right) = 1-\frac{\omega_p^2}{\omega^2},
\end{equation}

with the plasma frequency $\omega_p$. For the calculation of the interaction between molecular exciton and surface plasmon the dipole approximation is employed and it will be assumed that the molecular structure is close to the surface. Thus, the exponential decay term in equation (11) is approximated by one. 

The model which describes the interaction between $N$ molecules and a single quantized electromagnetic field mode is the Tavis-Cummings model \cite{PhysRev.170.379}. Thus, a generalization of the model for surface plasmon fields is employed in the following. The interaction Hamiltonian between the exciton molecules and the surface plasmon polaritons reads

\begin{equation}
    \hat{H}_\mathrm{i} = -\hat{\mu}\hat{\Vec{E}}_{SPP} = \hbar \sum_{i=1}^N g_i \left(\hat{a}\hat{\sigma}_+^{(i)} + \hat{a}^\dag\hat{\sigma}_-^{(i)}\right),
\end{equation}

where the coupling constants $g_i$ are given by

\begin{equation}
    g_i = - \sqrt{\frac{\hbar\omega(\Vec{k})}{2\epsilon_0 A}} \Vec{\mu}_{eg}^{i} \left(\Vec{u}_{\Vec{k}} + \mathrm{i} d \Vec{u}_z\right)
\end{equation}

with the transition dipole moments of the single molecules $\mu_{eg}^i$.

This is the final form of the coupling between surface plasmon polariton modes and excitonic structures, which will be utilized for the description of the plexciton dispersion of a PBI layer on a silver surface in the single excitation subspace. 
The plexciton Hamiltonian consists of the exciton Hamiltonian, eq. (1), the field Hamiltonian of the surface plasmon polariton mode $\hbar\omega(\vec{k})\hat{a}^\dag\hat{a}$, with the dispersion relation given by eq. (13), and the interaction Hamiltonian, eq. (15). Therefore, it is given by

\begin{equation}
    \hat{H} = \hat{H}_{\mathrm{exc}} + \hbar\omega(\Vec{k}) \hat{a}^\dag\hat{a} + \hat{H}_\mathrm{i}.
\end{equation}

A appropriate basis in the single excitation subspace for this Hamiltonian is obtained by taking the direct products of the cavity state $|1\rangle_c$ with the total ground state of the excitonic system and the cavity state $|0\rangle_c$ with the exciton basis states, eq. (4). The matrix representation of the plexciton Hamiltonian in this basis is given by

\begin{equation}
    H = 
    \begin{pmatrix}
    \hbar \omega(\Vec{k}) & \left[V\right] \\
    \left[V\right]^\dag & \left[H_{\mathrm{exc}}\right],
    \end{pmatrix}
\end{equation}

where $\hbar\omega(\vec{k})$ is the energy of the surface plasmon mode, $\left[H_{\mathrm{exc}}\right]$ is the matrix representation of the exciton Hamiltonian and $\left[V\right]$ is a row containing the couplings of the surface plasmon polariton field with the local excited states of the monomers, which are given by

\begin{equation}
    V_i = - \sqrt{\frac{\hbar\omega(\Vec{k})}{2\epsilon_0 A}} \Vec{\mu}_{eg}^{i} \left(\Vec{u}_{\Vec{k}} + \mathrm{i} d \Vec{u}_z\right).
\end{equation}

Diagonalisation of this matrix gives rise to the eigenenergies and eigenstates of the plexciton, which arise by the coupling of the exciton to a surface plasmon polariton mode with a given wave vector $\vec{k}$. For the calculation of plexciton absorption the spectral intensities $f_i$ of the eigenstates have to be calculated. These are given by the absolute square of the expansion coefficients $c_i$ of the basis state $|1\rangle_c\otimes|\phi_{0, \dots, 0, \dots, 0}\rangle$, representing an emitted photon and the total ground state of the molecules:

\begin{equation}
    f_i = |c_i|^2.
\end{equation}

\begin{figure}[H]
\includegraphics[width=0.9\textwidth]{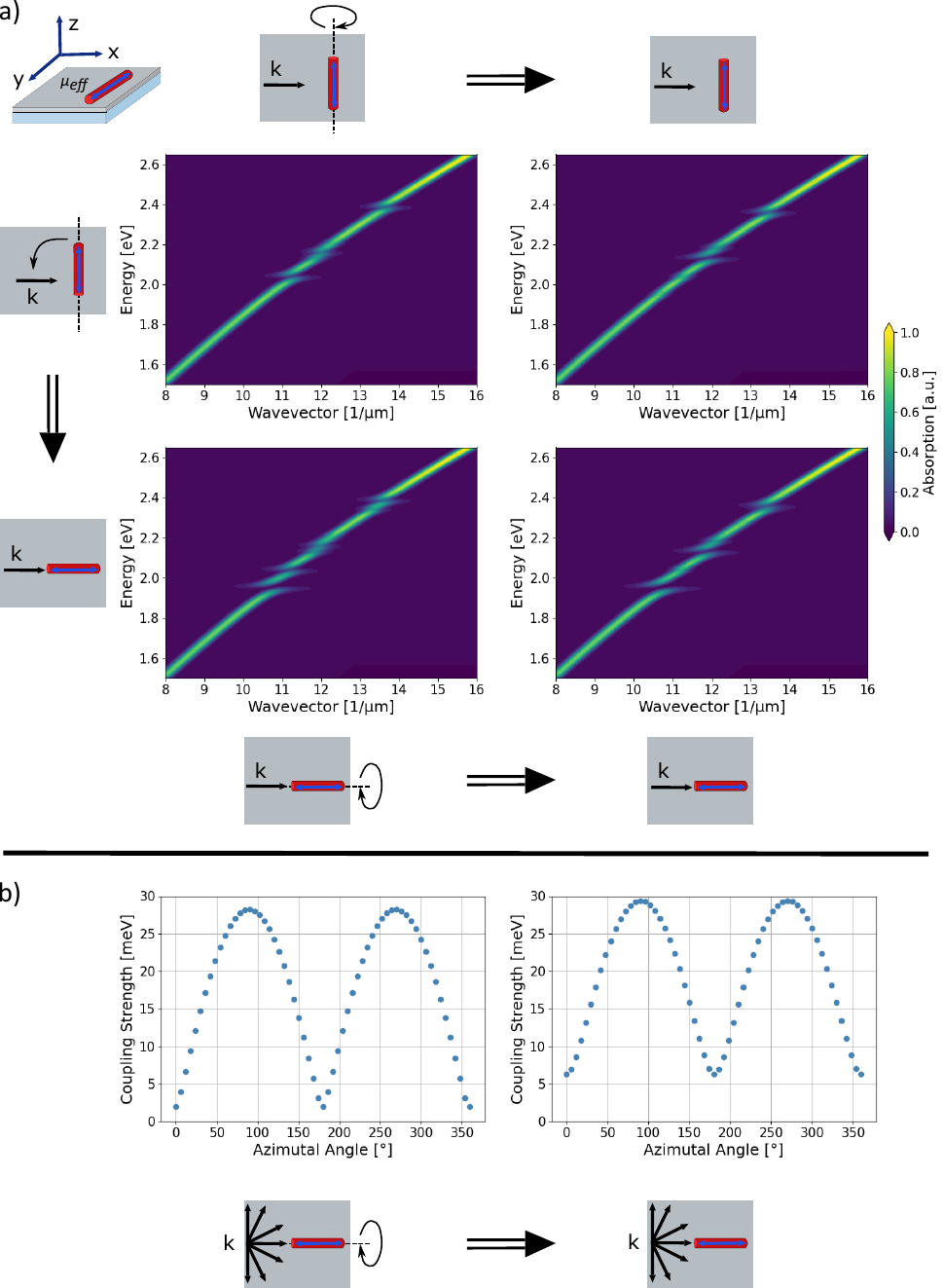}
\caption{a) Calculated plexciton absorption spectra for four different excitation geometrys. The top row has a perpendicular orientation of the transition dipole $\mu_{eff}$ and the plasmon wave vector k, while the lower row has a parallel arrangement of $\mu_{eff}$ and k, resulting in the low and the high coupling case , respectively. By going from the left to the right column the long helical strain axis is rotated by 90$^\circ$. b) Azimuthal angle-dependent coupling strength for the anisotropic anti-crossing for two different angle along the helical axis. \label{S_fig:Theoretical Anisotropic Plexciton Absorption} }
\end{figure}

Plexciton dispersion spectra are then simulated by calculating various plexciton absorption spectra with the excitons coupled to surface plasmon polariton modes with different magnitudes of the wave vector $|\Vec{k}|$. The absorption spectra are plotted against the magnitude of the wave vector, providing the dispersion spectra in figure \ref{S_fig:Theoretical Anisotropic Plexciton Absorption} for 4 different cases. We distinguish between the strong coupling where the wave vector $|\Vec{k}|$ is parallel to the transition dipole moment $\vec{\mu}_{eff}$ and the weak coupling where $|\Vec{k}|$ is perpendicular to $\vec{\mu}_{eff}$, corresponding to the lower and upper row, respectively. By rotating along the long strain axis of the helical aligned PBI-J-aggregates by 90$^\circ$ we go from the left column to the right column and vice versa. Note that in our experiments the absorption spectra/coupling strength should be an superposition of the upper and lower row for the two extreme cases, namely minimal and maximal coupling. Based on this, we have calculated the shown theoretical dispersions in the main text for strong and weak coupling from the mean values of the first and second column, respectively. The same applies to the calculated coupling strengths in figure \ref{S_fig:Theoretical Anisotropic Plexciton Absorption}b).

\subsection*{Computational Details}

Theoretical calculations for the description of the electronic structure, i.e. excitation energies and transition dipole moments, of the PBI monomers are carried out in the framework of time-dependent density functional theory (TD-DFT) with the hybrid $\omega$B97X-D3 functional \cite{lin2013long} in the Tamm-Dancoff approximation (TDA) as implemented in the orca software package \cite{neese2020orca} with def2-SVP basis set \cite{weigend2005balanced}. The couplings between the excited states of different monomers are calculated utilizing Mulliken transition charges, which were calculated with the Multiwfn program \cite{lu2012multiwfn}. Using these, the coupling is given by the Coulomb interaction between transition charges of excited states of different monomers, eq. (10). The coupling between the excited states of the molecules and surface plasmon polariton modes is calculated using equations (15) and (16). The outer substituents of the PBI are replaced by hydrogen atoms to reduce the computational demand. The couplings between neighbouring helices on the thin film are assumed to be negligible. Thus, all calculations are carried out considering a single helix. Constants in the plasmon dispersion, i.e. the dielectric constant of the PBI and the plasma frequency of silver, are chosen to reproduce the experimental fundamental dispersion relation of the surface plasmon polariton. Hereby the incidence angles given in the experimental data are connected to the in-plane wavevector $\Vec{k}$. The quantized area $A$ and the enhancement factor $d$ were chosen such that the calculated spectra approximately yield the experimental splittings of the anti-crossings. The calculation of the absorption spectrum of the PBI helix reveals that the theoretical maximum is located at around $2.37\ \mathrm{eV}$, while the experimental maximum is located at $1.95\ \mathrm{eV}$. To compensate for this energetic difference of $0.42\ \mathrm{eV}$, the theoretical surface plasmon polariton dispersion curve is accordingly shifted to higher energies. Then, the absorption spectra of the coupled system of exciton and surface plasmon polariton mode are calculated, and the result is shifted back by $0.42\ \mathrm{eV}$ to allow for comparison with the experimental data. For the calculation of the dispersion spectra it is assumed, that the helix is parallel to the metal surface. The coordinate system is chosen such that the z-axis is directed along the helix, the x-axis is perpendicular to the surface and the y-axis lies in the surface and is perpendicular to the helix. Notice that it is always the field component perpendicular to the surface that is enhanced. Two pairs of dispersion spectra are calculated. In the first pair, for a given helix orientation, two orientations of the field are applied, where the weak component is either perpendicular or parallel to the helix. In the second pair, the helix is rotated by $90^\circ$ about the z-axis. For both pairs, the orientations of the field give rise to a stronger and weaker coupling between the surface plasmon polariton mode and the excitonic states. The stronger (weaker) coupling manifests itself in a stronger (weaker) pronounced anti-crossing. Plexcitonic photoluminescence intensities were calculated as described in reference\cite{koponen2013absence}, i.e., we take the intensity to be given by the transition amplitudes of the plexcitonic state multiplied with the sum of the absolute squared Hopfield coefficients of the excitonic states in the plexcitonic wave function.

\subsection*{Model System for the Description of Photoluminescence Dispersion}

In the following, a model system for the description of the plexcitonic photoluminescence will be derived. Consider a relaxed molecular structure with a single electronic transition with transition frequency $\omega_m$, which is assumed to be already relaxed in its excited state. The structure is coupled to a surface plasmon polariton with the dispersion relation $\omega_c(k)$. The light-matter coupling strength is given by the coupling constant $g$, which is assumed to be independent of $k$. Thus, the Hamiltonian of the system is given by

\begin{equation}
    H = \hbar \begin{pmatrix}
     \omega_c(k) & g \\
    g & \omega_m
    \end{pmatrix}.
\end{equation}

The eigenenergies of this Hamiltonian are given by

\begin{equation}
    E_\pm = \frac{\hbar}{2} \left(\omega_c(k)+\omega_m \pm \sqrt{\delta^2+\Omega^2}\right),
\end{equation}

where we introduced the detuning frequency $\delta = \omega_c(k)-\omega_m$ and the new coupling constant $\Omega = 2g$.
The eigenstates to $E_\pm$ are given by

\begin{equation}
    \begin{aligned}
        &|+\rangle = \sin\theta |1,g\rangle + \cos\theta |0,e\rangle \\
        &|-\rangle = \cos\theta |1,g\rangle - \sin\theta |0,e\rangle,
    \end{aligned}
\end{equation}

with the state vector $|1,g\rangle$ ($|0,e\rangle$), which describes the state with the molecular structure in the ground (excited) state and one (no) photon in the surface plasmon polariton mode, and  the mixing angle $\theta$, which is defined by

\begin{equation}
    \tan(2\theta) = \frac{\Omega}{\delta}.
\end{equation}

For the calculation of the photoluminescence, we assume that the system relaxes into the diabatic states $|1,g\rangle$ and $|0,e\rangle$, i.e. photoluminescence occurs out of these states. Furthermore, we assume that the probability that the lower plexciton branch $|-\rangle$ relaxes into the diabatic state $|0(1), e(g)\rangle$ is given by $\left|\langle -|0(1),e(g)\rangle\right|^2$. For the upper polariton branch $|+\rangle$, we assume that it always relaxes into the molecular excited state $|0,e\rangle$ due to the ensemble of dark states in between lower and upper plexciton branch, which arise by the coupling of multiple helices to the surface plasmon polariton mode. It is in question, if this relaxation only occurs for the molecular part of the upper plexciton branch, i.e. the probability that photoluminescence takes place from the molecular excited state after relaxation from the upper plexciton branch is proportional to $\left|\langle +|0,e\rangle\right|^2$. In the following this case will be denoted as a). Otherwise, we can assume that the upper polariton branch always relaxes into the molecular excited state, i.e. photoluminescence from the excited molecular state always takes place after relaxation from the upper plexciton branch. This case will be denoted as b). For both cases the average energy of the emitted photons $E_{\textbf{fl}}(k)$ is given by the average photon energy of the emitted photons divided by the probability that an absorbed photon is also emitted. For case a) this yields the expression

\begin{equation}
    \begin{aligned}
        E_{\text{fl}}^a(k) &= \hbar\frac{\cos^2\theta\left(\cos^2\theta\omega_c(k)+\sin^2\theta\omega_m\right) + \sin^2\theta\cos^2\theta\omega_m}{\cos^2\theta+ \cos^2\theta\sin^2\theta} \\
        &= \frac{\hbar}{3-\cos(2\theta)}\left( \omega_c(k) + 2\omega_m + \cos(2\theta)\left(\omega_c(k)-2\omega_m\right) \right)
    \end{aligned}
\end{equation}

and for case b) it yields

\begin{equation}
    \begin{aligned}
    E_{\text{fl}}^b(k) &= \hbar \left( \cos^2\theta \left( \cos^2\theta\omega_c(k)+\sin^2\theta\omega_m \right) + \sin^2\theta \omega_m\right)\\
    &= \hbar \left(\cos^4\theta\omega_c(k) + \sin^2\theta \left( 1 + \cos^2\theta\right) \omega_m\right).
    \end{aligned}
\end{equation}

\begin{figure}[H]
\includegraphics[width=0.9\textwidth]{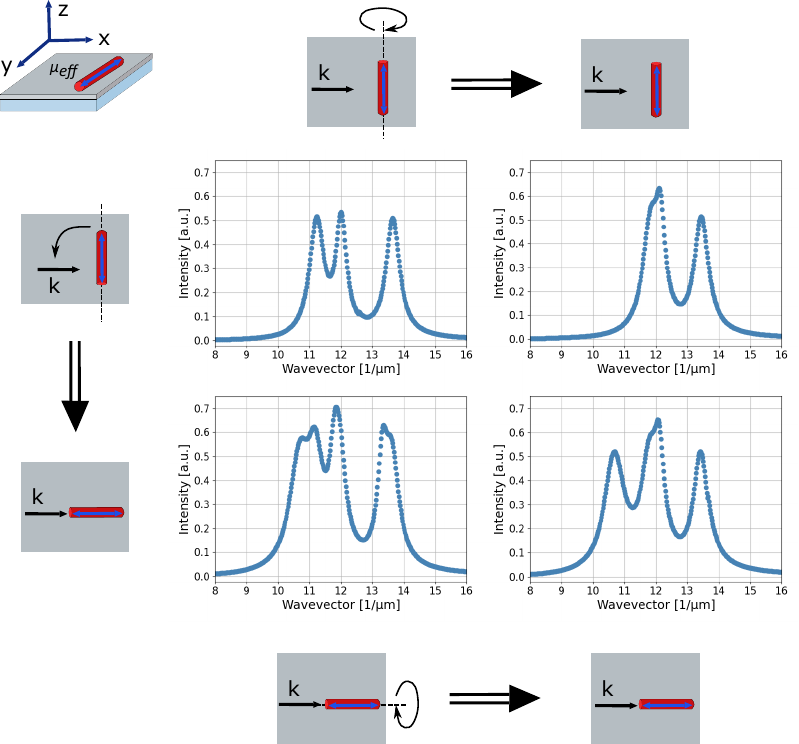}
\caption{Calculated plexciton photoluminescence intensity for four different excitation geometrys analogue to figure \ref{S_fig:Theoretical Anisotropic Plexciton Absorption}. The top row has a perpendicular orientation of the transition dipole $\mu_{eff}$ and the plasmon wave vector k, while the lower row has a parallel arrangement of $\mu_{eff}$ and k, resulting in the low and the high coupling case , respectively. By going from the left to the right column the long helical strain axis is rotated by 90$^\circ$. \label{S_fig:Theoretical Anisotropic Photoluminescence}}
\end{figure}

For the calculation of these quantities, it was assumed that the dispersion of the surface plasmon polariton is linear. The linear dispersion was fitted to the experimental dispersion in the region, where the detuning to the molecular excitation is high, and therefore the two systems are not coupled. The other constants were chosen according to the experimental values, i.e. $ \hbar g=0.027\ \text{eV}$ and $\hbar \omega_m = 1.849\ \text{eV}$. The photoluminescence intensity for the same four geometries is shown in figure \ref{S_fig:Theoretical Anisotropic Photoluminescence} while the dispersion is shown in the main text.

\newpage
\bibliography{references}